\documentclass[preprint,prd,aps,showpacs,superscriptaddress,preprintnumbers,nofootinbib,floatfix]{revtex4}

\def\DLB{\Delta L\!B}

\usepackage{amsmath}
\usepackage{amssymb}
\usepackage{bm}% bold math
\usepackage{dcolumn}
\usepackage{graphicx}
\usepackage{enumerate}
\usepackage{longtable}
\begin{document}
%----------------------------------------------------------------

%================================================================
%\date{\today}
%================================================================

%\preprint{}

%================================================================
\title{%
Tenth-Order Electron Anomalous Magnetic Moment ---
Contribution of Diagrams without Closed Lepton Loops
}

%----------------------------------------------------------------
\affiliation{Kobayashi-Maskawa Institute for the Origin of Particles and the Universe (KMI), Nagoya University, Nagoya, 464-8602, Japan}
\affiliation{Department of Physics, Nagoya University, Nagoya, 464-8602, Japan }
\affiliation{Laboratory for Elementary Particle Physics, Cornell University, Ithaca, NY, 14853, U.S.A. }
\affiliation{Amherst Center for Fundamental Interactions,
Department of Physics, University of Massachusetts,
Amherst, MA, 01003, U.S.A.}
\affiliation{Nishina Center, RIKEN, Wako, 351-0198, Japan }
%----------------------------------------------------------------

\author{Tatsumi Aoyama}
\affiliation{Kobayashi-Maskawa Institute for the Origin of Particles and the Universe (KMI), Nagoya University, Nagoya, 464-8602, Japan}
\affiliation{Nishina Center, RIKEN, Wako, 351-0198, Japan }

\author{Masashi Hayakawa}
\affiliation{Department of Physics, Nagoya University, Nagoya, 464-8602, Japan }
\affiliation{Nishina Center, RIKEN, Wako, 351-0198, Japan }

\author{Toichiro Kinoshita}
\affiliation{Laboratory for Elementary Particle Physics, Cornell University, Ithaca, NY, 14853, U.S.A. }
\affiliation{Amherst Center for Fundamental Interactions,
Department of Physics, University of Massachusetts,
Amherst, MA, 01003, U.S.A.}
\affiliation{Nishina Center, RIKEN, Wako, 351-0198, Japan }

\author{Makiko Nio}
\affiliation{Nishina Center, RIKEN, Wako, 351-0198, Japan }

%================================================================
\begin{abstract}
This paper presents a detailed account of evaluation
of the electron anomalous magnetic moment $a_e$ 
which arises from the gauge-invariant set, called Set~V, consisting of
6354 tenth-order Feynman diagrams without
closed lepton loops.
The latest value of the 
sum of Set~V diagrams evaluated by the Monte-Carlo integration  
routine VEGAS is 8.726~(336)$(\alpha/\pi)^5$,
which replaces the very preliminary value reported in 2012.
Combining it with other 6318 tenth-order diagrams published previously
we obtain 7.795~(336)$(\alpha/\pi)^5$ as the complete mass-independent
tenth-order term. 
Together with the improved value of the eighth-order term this 
leads to $a_e(\text{theory}) = 1~159~652~181.643~(25)(23)(16)(763) \times 10^{-12}$,
where first three uncertainties are from the eighth-order term,
tenth-order term, and hadronic and elecroweak terms.
The fourth and largest uncertainty is from 
$\alpha^{-1} = 137.035~999~049~(90)$,
the fine-structure constant derived from the rubidium recoil measurement.
Thus, 
$a_e(\text{experiment}) - a_e(\text{theory}) = -0.91~(0.82) \times 10^{-12}$. 
Assuming the validity of the standard model, we obtain the fine-structure constant
$\alpha^{-1}(a_e) = 137.035~999~1570~(29)(27)(18)(331)$,
where uncertainties are from the eighth-order term, tenth-order term, hadronic and electroweak terms, and the measurement of $a_e$.
This is the most precise value of $\alpha$
available at present and 
provides a stringent constraint on possible
theories beyond the standard model. 
\end{abstract}
%================================================================

%----------------------------------------------------------------
%

\maketitle
%\tableofcontents

%================================================================
\section{Introduction and Summary}
\label{sec:intro}

The anomalous magnetic moment of the electron, $a_e \equiv (g-2)/2$, 
has played an important role in testing the validity of 
quantum electrodynamics (QED) and the standard model of particle physics. 
The latest measurement of $a_e$ by the Harvard group has reached the precision
of $0.24 \times 10^{-9}$ 
\cite{Hanneke:2008tm,Hanneke:2010au}:
\begin{eqnarray}
  a_e(\text{HV08})= 1~159~652~180.73~(0.28) \times 10^{-12} \quad [0.24\text{ppb}]~.
\label{a_eHV08}
\end{eqnarray}
A new apparatus for measuring $g-2$ of electron and positron
with even higher precision is being constructed by the same group
\cite{Fogwell-Hoogerheide:2013thesis}.
In order to test QED to such a precision
it is necessary to have a theoretical value of
 the tenth-order term since
\begin{equation}
  (\alpha/\pi)^5 \sim 0.07 \times 10^{-12},
\end{equation}
where $\alpha$ is the fine-structure constant.

In the standard model the contribution to $a_e$ comes from
three types of interactions, electromagnetic, hadronic,
and electroweak, which may be written as
\begin{equation}
  a_e = a_e (\text{QED}) + a_e (\text{hadronic}) + a_e (\text{electroweak}),
\label{aeSM}
\end{equation}
although $a_e (\text{hadronic})$ contains contribution of 
electromagnetic interaction 
and $a_e (\text{electroweak})$ contains contribution of
electromagnetic and hadronic interactions in higher orders.
In the framework of the standard model 
the dominant contribution comes from $a_e (\text{QED})$.
$a_e (\text{hadronic})$ and $a_e (\text{electroweak})$ 
provide only small corrections.
However the latter two cannot be ignored in comparing theory with measurements.

The QED contribution can be evaluated by the perturbative expansion
in $\alpha/\pi$:
\begin{equation}
  a_e (\text{QED}) = \sum_{n=1}^{\infty} \left ( \frac{\alpha}{\pi} \right )^n a_e^{(2n)}, 
\label{aeQED}
\end{equation}
where $a_e^{(2n)}$ is finite due to the renormalizability of QED
and may be written in general as
\begin{eqnarray}
  a_e^{(2n)}
  = A_1^{(2n)} 
  + A_2^{(2n)} (m_e/m_\mu)
  + A_2^{(2n)} (m_e/m_\tau) 
  + A_3^{(2n)} (m_e/m_\mu, m_e/m_\tau)
\label{ae2n}
\end{eqnarray}  
to exhibit the dependence on the muon mass and tau-particle masses.
We use the electron-muon mass ratio
$m_e/m_\mu = 4.836~331~66~(12) \times 10^{-3}$
and the electron-tau mass ratio
$m_e/m_\tau = 2.875~92~(26) \times 10^{-4}$
\cite{Mohr:2012tt}.

The first three terms of $A_1^{(2n)}$ are known analytically
\cite{Schwinger:1948iu,Petermann:1957,Sommerfield:1958,Laporta:1996mq}.
Their numerical values are 
\begin{eqnarray}
  A_1^{(2)} &=& 0.5,  \nonumber \\ 
  A_1^{(4)} &=& -0.328~478~965~579~193 \ldots,  \nonumber \\ 
  A_1^{(6)} &=&  1.181~241~456 \ldots.  
\label{A1_246}
\end{eqnarray}  
The value of $A_1^{(8)}$,
which has contributions from 891 Feynman diagrams,
 is obtained mostly by numerical integration \cite{ae10:PRL}.
It is being improved continually by further numerical work.
The latest value
\begin{eqnarray}
A_1^{(8)} &=& -1.912~98~(84),
\label{A1_8}  
\end{eqnarray}  
obtained by a substantial increase in the sampling statistics
of VEGAS \cite{Lepage:1977sw} calculation,
is a factor 2.4 improvement over the published result
\cite{ae10:PRL}.

The term $A_1^{(10)}$ has contributions from 
12,672 vertex diagrams, which may be classified
into six gauge-invariant sets,
further subdivided into 32 gauge-invariant subsets depending
on the type of lepton loop subdiagrams.
Thus far, the results of numerical evaluation of 31 gauge-invariant subsets,
which consists of 6318 vertex diagrams, have been published 
\cite{Kinoshita:2005sm,Aoyama:2008gy,Aoyama:2008hz,Aoyama:2010yt,Aoyama:2010pk,Aoyama:2010zp,Aoyama:2011rm,Aoyama:2011zy,Aoyama:2011dy,Aoyama:2012fc}.
The results of all 10 subsets of Set~I, consisting of 208 vertex diagrams, 
have been confirmed by Ref.~\cite{Baikov:2013ula}.
All these diagrams have closed lepton loops and thus contribute
also to $A_2^{(10)}$ and/or $A_3^{(10)}$. 

The remaining set, called Set~V, 
consists of 6354 Feynman diagrams which
do not have closed lepton loops (denoted as {\it q}-type diagrams).
It is the largest and most difficult one to evaluate.
This paper presents a detailed account of evaluation of Set~V diagrams, 
and gives the latest numerical value.
The presented value is more accurate and reliable than the preliminary 
one reported in Ref.~\cite{ae10:PRL}, 
not only because of the increase of the statistics of Monte-Carlo integration, 
but also by the incorporation of the qualitative improvements explained 
in Section~\ref{sec:result}. 

Integrals of the Set~V are huge and complicated
so that their evaluation requires an enormous amount of work.
A systematic and fully automatic approach is an absolute necessity 
to carry out such a project.
To meet this challenge we have developed an algorithm and its implementation, 
{\sc gencode}{\it N} \cite{Aoyama:2005kf, Aoyama:2007bs}, 
which automatically converts the diagrammatic information, 
specifying how virtual photon lines are attached to the lepton lines,
into a FORTRAN code free from ultraviolet and infrared divergences.

The evaluation of the tenth-order diagrams boils down to the numerical 
integration 
on a 13-dimensional unit cube 
onto which a hyperplane of 14 Feynman parameters is mapped. 
The integrals are evaluated 
by the adaptive-iterative 
Monte-Carlo integration routine VEGAS \cite{Lepage:1977sw}. 
For this calculation, 
the RIKEN Supercomputing Systems RSCC and RICC are used intensively. 
The results are summarized in Table~\ref{table:X001}. 
Auxiliary quantities required for restoring the standard on-shell renormalization are
listed in  Table~\ref{Table:residual_const}.
From these Tables we obtain
\begin{equation}
A_1^{(10)} [\text{Set~V}] = 8.726~(336).
\label{eq:ae10th_set5}
\end{equation}
Adding this to the values of the other 31 gauge-invariant sets,
which were evaluated and published previously
\cite{Kinoshita:2005sm,Aoyama:2008gy,Aoyama:2008hz,Aoyama:2010yt,Aoyama:2010pk,Aoyama:2010zp,Aoyama:2011rm,Aoyama:2011zy,Aoyama:2011dy,Aoyama:2012fc},
we now have an improved value of the sum of all 12,672 diagrams of tenth-order
\begin{equation}
A_1^{(10)} = 7.795~(336),
\label{A1_10}
\end{equation}
which replaces the very preliminary value reported in Ref.~\cite{ae10:PRL}.

The mass-dependent terms $A_2$ and $A_3$ of the fourth and sixth
orders are known 
\cite{Elend:1966a,Samuel:1990qf,Li:1992xf,Laporta:1992pa,Laporta:1993ju,Passera:2006gc},
\begin{eqnarray}
  A_2^{(4)}(m_e/m_\mu)   &=& 5.197~386~67~(26) \times 10^{-7},  \nonumber \\ 
  A_2^{(4)}(m_e/m_\tau)  &=& 1.837~98~(34) \times 10^{-9},  \nonumber \\ 
  A_2^{(6)}(m_e/m_\mu)   &=& -7.373~941~55~(27) \times 10^{-6},  \nonumber \\ 
  A_2^{(6)}(m_e/m_\tau)  &=& -6.583~0~(11) \times 10^{-8},  \nonumber \\ 
  A_3^{(6)}(m_e/m_\mu, m_e/m_\tau) &=&  1.909~(1) \times 10^{-13},
\label{A23_46}
\end{eqnarray}  
and those of eighth and tenth order terms can be found in Refs.~%
\cite{Kinoshita:2005sm,Aoyama:2008gy,Aoyama:2008hz,Aoyama:2010yt,Aoyama:2010pk,Aoyama:2010zp,Aoyama:2011rm,Aoyama:2011zy,Aoyama:2011dy,Aoyama:2012fc,Kurz:2013exa}
\begin{eqnarray}
A_2^{(8)}(m_e/m_\mu)  &=& 9.161~970~703~(373) \times 10^{-4},  \nonumber \\ 
A_2^{(8)}(m_e/m_\tau)  &=& 7.429~24~(118) \times 10^{-6},  \nonumber \\ 
A_3^{(8)}(m_e/m_\mu, m_e/m_\tau) &=&  7.4687~(28) \times 10^{-7}, \nonumber \\
A_2^{(10)}(m_e/m_\mu)  &=& -0.003~82~(39) .  
\label{A23_810}
\end{eqnarray}  
Our evaluation of $A_2^{(8)}(m_e/m_\mu)$ and $A_2^{(8)}(m_e/m_\tau)$ has 
been confirmed by the analytic calculations of Refs.~\cite{Kurz:2013exa,Kataev:2012kn}.\footnote{%
There is a typo in Table~I of Ref.~\cite{ae10:PRL} for the contribution 
from Group~I(d) to $A_2^{(8)}(m_e/m_\tau)$ in which the actual value is 
$0.8744(1) \times 10^{-8}$, as pointed out in Ref.~\cite{Kurz:2013exa}.
}

Recently, the possible non-perturbative effect of QED 
to the order of $\alpha^5$ of the electron $g-2$ 
was pointed out \cite{Mishima:2013ama,Fael:2014nha}, 
but shown to be absent \cite{Melnikov:2014lwa,Fael:2014nha,Eides:2014swa} 
in accord with the earlier studies 
of Refs.~\cite{BraunM.A.:1968xia,Barbieri:1973lza} 
applied to the electron $g-2$. 
Ref.~\cite{Hayakawa:2014tla} presents an approach different from those 
of Refs.~\cite{Melnikov:2014lwa,Fael:2014nha,Eides:2014swa}.

The latest values of the leading order, 
next-to-leading order (NLO),
next-to-next-to-leading order (NNLO) contributions
of the hadronic vacuum-polarization (v.p.) are given in Refs.~\cite{Nomura:2012sb,Kurz:2014wya} 
\begin{eqnarray}
  a_e (\text{had.~v.p.}) 
  &=& 1.866~(10)_\text{exp}~(5)_\text{rad} \times 10^{-12} , \nonumber \\
  a_e (\text{NLO had.~v.p.}) 
  &=& -0.2234~(12)_\text{exp}~(7)_\text{rad} \times 10^{-12} , \nonumber \\
  a_e (\text{NNLO had.~v.p.}) 
  &=& 0.028~(1) \times 10^{-12} , 
\label{hadvp}
\end{eqnarray}  
and the hadronic light-by-light-scattering (\textit{l-l}) term is given in Ref.~\cite{Prades:2009tw}:
\begin{equation}
  a_e (\text{had.~\textit{l-l}}) = 0.035~(10) \times 10^{-12} .
\label{hadll}
\end{equation}
The electroweak contribution has been obtained from the analytic form of
the one-loop \cite{Fujikawa:1972fe}
and two-loop \cite{Czarnecki:1996ww,Knecht:2002hr,Czarnecki:2002nt}
electroweak effects on the muon $g-2$, 
adapted for the electron. 
We quote the value summarized and updated in Ref.~\cite{Mohr:2012tt}:
\begin{equation}
a_e (\text{electroweak}) = 0.0297~(5) \times 10^{-12} .
\label{electroweak}
\end{equation}

To compare the theoretical prediction with the measurement (\ref{a_eHV08}),
we need the value of the fine-structure constant $\alpha$ determined
by a method independent of $g-2$.
The best $\alpha$ available at present is the one derived 
from the precise value of $h/m_\text{Rb}$, 
which is obtained by the measurement of the recoil velocity of Rubidium atoms 
on an optical lattice \cite{Bouchendira:2010es},
combined with the very precisely known Rydberg constant and $m_\text{Rb}/m_e$
\cite{Mohr:2012tt}:
\begin{equation}
\alpha^{-1} (\text{Rb10}) = 137.035~999~049~(90)~~~[0.66~\text{ppb}].
\label{eq:alpha}
\end{equation}
With this $\alpha$ the theoretical prediction of $a_e$ becomes
\begin{equation}
a_e (\text{theory}) = 1~159~652~181.643~(25)(23)(16)(763) \times 10^{-12},
\label{eq:aetheory}
\end{equation}
where the first, second, third, and fourth uncertainties come
from the eighth-order term (\ref{A1_8}), the tenth-order term (\ref{A1_10}),
the hadronic (\ref{hadvp})--(\ref{hadll}) and electroweak (\ref{electroweak}) corrections, 
and the fine-structure constant (\ref{eq:alpha}), respectively.
This is in good agreement with the experiment (\ref{a_eHV08}):
\begin{equation}
a_e (\text{HV08}) - a_e (\text{theory}) = -0.91~(0.82) \times 10^{-12}.
\label{eq:aediff}
\end{equation}

The intrinsic theoretical uncertainty ($\sim 38 \times 10^{-15}$)
of $a_e (\text{theory})$ is less than 1/20 of the uncertainty due to 
the fine-structure constant (\ref{eq:alpha}).
This means that a more precise value of $\alpha$ than Eq.~(\ref{eq:alpha}) 
can be obtained assuming that 
QED and the standard model are valid and solving 
the equation $a_e (\text{theory}) = a_e (\text{experiment})$ for $\alpha$: 
\begin{equation}
\alpha^{-1} (a_e) = 137.035~999~1570~(29)(27)(18)(331)~~~[0.25~\text{ppb}],
\label{eq:alpha_ae}
\end{equation}
where the uncertainties are from
the QED terms (\ref{A1_8}), (\ref{A1_10}), combined 
hadronic (\ref{hadvp}), (\ref{hadll}) and electroweak (\ref{electroweak}) terms,
and the measurement (\ref{a_eHV08}) of $a_e (\text{HV08})$, in that order.
This provides a stringent constraint on possible
theories beyond the standard model. 
It can be made even more stringent
by improved measurements of $a_e$.

Section~\ref{preliminaryremarks}
describes how we organized diagrams of Set~V
into a smaller number of independent integrals.
Section~\ref{sec:codegeneration} describes the steps involved in
the automatic code generation by {\sc gencode}{\it N}. 
Section~\ref{sec:result} discusses computational problems encountered 
in the numerical integration.
Section~\ref{sec:discussion} is devoted to the discussion of some 
technical problems encountered in our work. 

Appendix~\ref{sec:appendixX253} describes how
{\it K}-operation, {\it R}-subtraction and {\it I}-operation 
\cite{Cvitanovic:1974sv,Aoyama:2005kf,Aoyama:2007bs}
introduced in Section~\ref{sec:codegeneration}
work, choosing the diagram {\rm X253} as an example.
Actually {\rm X253} is one of the exceptional diagrams 
(the other is {\rm X256})
for which the implementation of {\it I}-operation 
in {\sc gencode}{\it N} requires slight modification 
according to the definition of the residual part of vertex renormalization 
constant with insertion of a two-point vertex $L_{n^*}^{\rm R}$, 
which has been treated manually. 
This is manifested first at these two diagrams in tenth order, 
while it is absent in the eighth and lower order diagrams. 
Thus, the evaluation of the eighth order diagrams (called Group~V) 
that relies on {\sc gencode}{\it N} is correct. 
The detail will be fully discussed.
Appendix~\ref{sec:appendixRes} describes our approach to summing up of
the residual renormalization terms of Set~V.

%================================================================
\section{Reducing the number of integrals}
\label{preliminaryremarks}

Our evaluation of the tenth-order diagrams of Set~V relies on the 
numerical integration. 
The combined uncertainty $\sigma_N$ of $N$ independent integrals
grows roughly as $\sqrt N$.
Thus $\sigma_N$ becomes large for large $N$ even if each integral
has a small uncertainty.
This can be a particularly big headache for the Set~V, for which $N=6354$.

\begin{figure*}[tb]
\includegraphics[width=\textwidth]{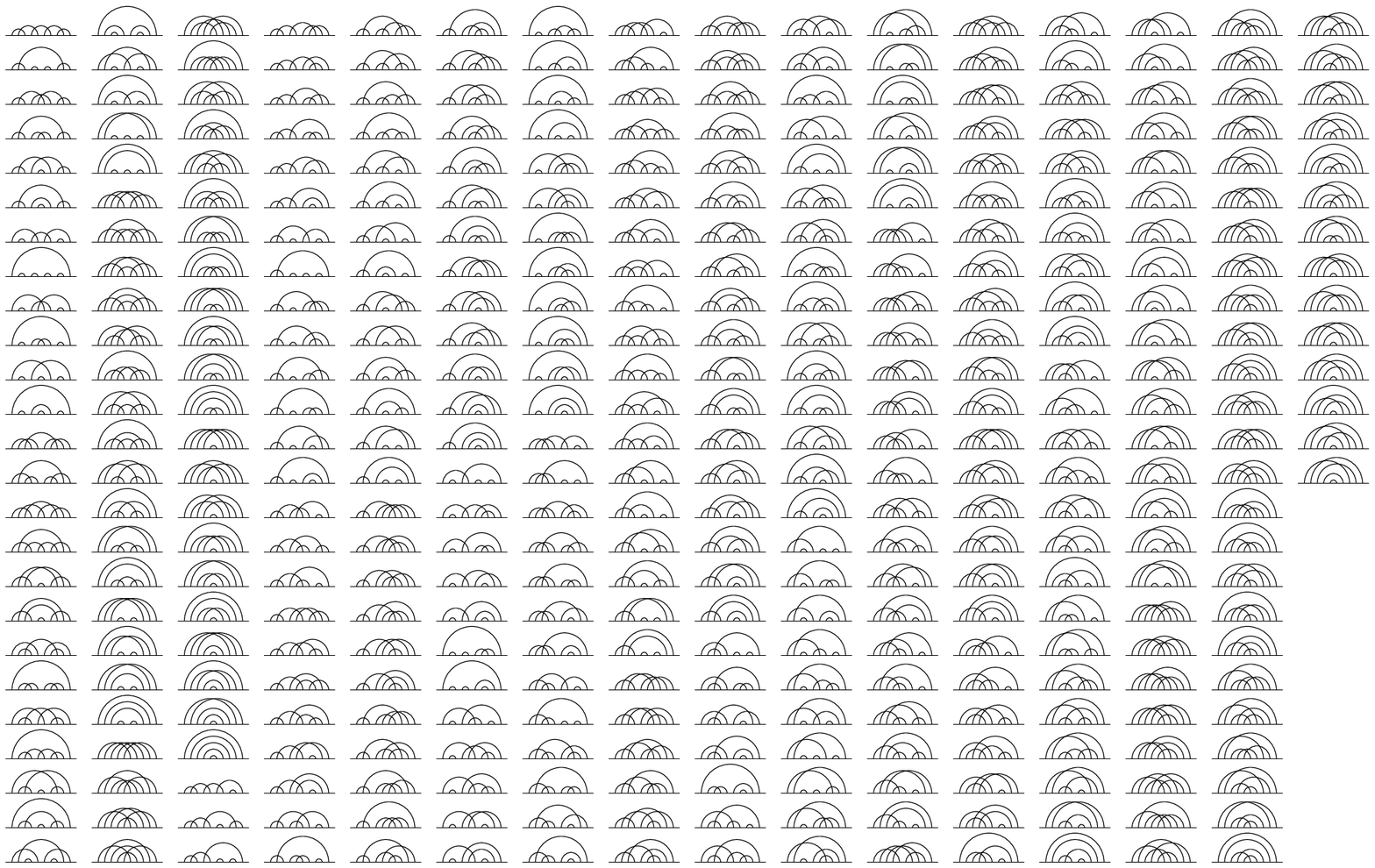}
\caption{
\label{fig:M10}
Overview of 389 diagrams which represents 6354 vertex diagrams of Set~V.
The horizontal solid lines represent the electron propagators
in a constant weak magnetic field.
Semi-circles stand for photon propagators.
The left-most figures are denoted as {\rm X001}--{\rm X025} 
from the top to the bottom.
The top figure in the second column from the left is denoted {\rm X026}, 
and so on.}
\end{figure*}

It is thus important to reduce the number of independent
integrals as much as possible.
For this purpose the technique based on the Ward-Takahashi identity
developed previously \cite{Cvitanovic:1974sv} is found to be handy.
It is based on the observation that a set of nine vertex diagrams,
which are derived from the self-energy-like diagram
${\cal G}$ of Fig.~\ref{fig:M10}
as the coefficients of terms linear in the external magnetic field,
share features which enable us to combine them
into a single integral.
Let $\Lambda^{\nu} (p, q)$ be the sum of these nine vertex diagrams,
where $p-q/2$ and $p+q/2$ are the 4-momenta of incoming and outgoing
lepton lines and $(p-q/2)^2 =(p+q/2)^2=m^2$.
The number of such sums is $6354/9 = 706$.
By taking time-reversal symmetry into account, 
the total number of independent integrals is reduced further to 389.
This is still large but far more manageable.

Let $\Sigma (p)$ be the integral representing 
the self-energy part of a diagram ${\cal G}$ of Fig.~\ref{fig:M10} 
(namely the part independent of the magnetic field). 
With the help of the Ward-Takahashi identity, we can rewrite 
$\Lambda^{\nu} (p, q)$  as
\begin{equation}
\Lambda^\nu (p,q) \simeq - q_\mu \left [ \frac{\partial \Lambda_\mu (p,q)}{\partial q_\nu} \right ]_{q=0} - \frac{\partial \Sigma (p)}{\partial p_\nu}
\label{WTderived}
\end{equation}
in the small $q$ limit. 
The $g-2$ term $M_{\cal G}$ is projected out from 
either the left-hand side ({\it lhs}) or
the right-hand side ({\it rhs}) of Eq.~(\ref{WTderived}).
Considerable numerical cancellation is expected among the nine terms 
on the {\it lhs} of Eq.~(\ref{WTderived}).
In fact the {\it rhs} exhibits
the consequence of such a cancellation at the algebraic level.
Thus  starting from the {\it rhs}
enables us to reduce the amount of computing time
substantially (by at least a factor 5), 
and also to improve the precision of numerical results significantly.

Since these integrals have UV-divergent subdiagrams,
they must be regularized by some means.
For the diagrams of Set~V the Feynman cut-off,
which is a sort of ``mass'' of virtual photons, 
works fine as the regulator.
Suppose all integrals, including renormalization terms,
are initially regularized by the Feynman cut-off.
Of course the final renormalized result
is finite and well-defined in the limit of infinite cut-off mass.

%================================================================
\section{Formulation}
\label{sec:codegeneration}

Most of these diagrams are so huge and complicated that
numerical integration is the only viable option at present.
In order to evaluate them on a computer which requires that
every step of computation is finite, however, it is necessary to remove
all sources of divergence of the integrand {\it before}
 carrying out integration.
This is achieved by the introduction of {\it K}-operation
which deals with the UV divergences
\cite{Cvitanovic:1974sv,Aoyama:2005kf},  
and {\it R}-subtraction and {\it I}-operation
which deal with the IR divergences
\cite{Cvitanovic:1974sv,Aoyama:2007bs}.  
See \ref{subsec:UVsubtractionterms} and \ref{subsec:IRsubtractionterms}
for more details.

In practice it is very difficult
to carry out such a calculation without making mistakes
because of the gigantic size of the integral
and a large number of terms
required for renormalization. 
To deal with this problem we developed an automatic code-generating
algorithm {\sc gencode}{\it N} \cite{Aoyama:2005kf,Aoyama:2007bs},
 in which {\it N} implies that it works for the {\it q}-type diagrams
of any order {\it N} in the perturbation theory of QED.

%================================================================
\subsection{Diagram Generation}
\label{subsec:diagramgeneration}

The Feynman diagrams of Set~V have the structure that ten vertices 
along the electron line are 
connected by the virtual photons $a$, $b$, $c$, $d$, $e$, 
and thus specified by the pairing patterns of how these 
vertices are connected. 
Excluding patterns which are not one-particle irreducible, 
and taking time-reversal invariance into account, 
we obtain 389 different patterns which are represented by 
the diagrams of Fig.~\ref{fig:M10}. 
They are named as ${\rm X}{\it nnn}$, ${\it nnn} = 001, 002, \ldots, 389$.

The diagram $\rm X{001}$ represents the diagram at the upper left corner 
of Fig.~\ref{fig:M10}.
Subsequent expressions represent diagrams placed below $\rm X{001}$ until $\rm X{025}$,
and $\rm X{026}$ corresponds to the diagram placed to the right of $\rm X{001}$, and so on.
Diagrams $\rm X{001}$ to $\rm X{072}$ are time-reversal symmetric
and diagrams $\rm X{073}$ to $\rm X{389}$ are asymmetric.  
Within each group they are arranged in a lexicographical order. 

Each of these patterns can be expressed by a ``one-line'' statement of 
an ordered sequence of ten vertices labelled by the attached photon line 
indices. 
The diagram $\rm X{001}$ is identified by the sequence of paring pattern 
{\it abacbdcede}, which means that 
the first and third vertices from the left end of the electron line
are connected by the virtual photon $a$, the second and fifth vertices
are connected by the virtual photon $b$, and so on.

All Feynman diagrams of Set~V 
have a common feature except for the pattern of pairing of vertices. 
The integral of Set~V can thus be generated from a single 
master code by providing the simple diagram-specific information
${\rm X}{\it nnn}$ as the input.
It is important to 
note that this pairing pattern 
not only specifies completely the structure of the unrenormalized diagram 
but also represents the structure of all UV-divergent and IR-divergent 
subdiagrams required for renormalization. 

%================================================================
\subsection{Construction of unrenormalized integral}
\label{subsec:unrenormalizedintegral}

From a one-line statement specifying the diagram 
${\cal G} \equiv {\rm X}{\it nnn}$, 
the {\it rhs} of Eq.~(\ref{WTderived}) is translated into a
momentum space integral applying the Feynman-Dyson rule
(assuming the Feynman cutoff).
Introducing Feynman parameters $z_1$, $z_2$, \dots, $z_9$ for the electron
propagators,
where $z_i$ is for the $i$-th electron line from the left end of the diagram,
 and $z_a$, $z_b$, $z_c$, $z_d$, $z_e$ for the photon propagators,
it carries out the momentum integration {\it exactly} using
a home-made table written 
in FORM \cite{Vermaseren:2000nd}.
This leads to an integral of the form
\begin{equation}
M_{\cal G} = \left ( \frac{-1}{4} \right )^5 4 !
\int (dz)_{\cal G} \left [ \frac{1}{4} \left ( \frac{E_0 + C_0}{U^2V^4}
    + \frac{E_1+C_1}{U^3 V^3} + \ldots \right )
    + \left ( \frac{N_0+Z_0}{U^2V^5} + \frac{N_1+Z_1}{U^3V^4} + \ldots \right )
      \right ],
\label{M_G}
\end{equation}
where $E_n, C_n, N_n$ and $Z_n$ are functions of Feynman parameters $z_i$
and ``symbolic'' building blocks $A_i, B_{ij}, C_{ij}$ for
electron lines $i, j = 1, 2, \ldots, 9$.
$n$ is the number of {\it contractions} 
(see Refs.~\cite{Cvitanovic:1974sv,Kinoshita:1990} for definitions).
See, for example, Ref.~\cite{Aoyama:2005kf} for definitions of $B_{ij}$ and $C_{ij}$.
$U$ is the Jacobian of transformation from the momentum variables
to Feynman parameters. $(dz)_{\cal G}$ is defined by
\begin{equation}
(dz)_{\cal G} = \prod_{k \in {\cal G}} dz_k\ \delta \left( 1 - \sum_{k \in {\cal G}} z_k \right).
\label{def:dz}
\end{equation}
$A_i$ is the {\it scalar current}, which satisfies an analogue
of Kirchhoff's laws for electric current, and has the form
\begin{equation}
A_i =  \frac{1}{U} \sum_{j=1}^9 ( \delta_{ij} U - z_j B_{ij} ).
\end{equation}
$V$ is obtained by combining all denominators of
propagators into one with the help of the
Feynman parameters.  It has a form common to all diagrams of Fig.~\ref{fig:M10}:
\begin{equation}
V= \sum_{i=1}^9 z_i (1-A_i) m^2 + \sum_{\kappa=a}^e z_\kappa \lambda^2,
\label{def:V}
\end{equation}
where $m$ and $\lambda$ are rest masses of electron and photon,
respectively.
Of course, $\lambda$ must be sent to 0 in the end.

The explicit forms of $U$ and $B_{ij}$ as functions of Feynman parameters 
depend on the structure of the diagram ${\cal G}$. 
Once they are determined, $A_i$ and $V$ have common expression
for all diagrams of Set~V.
The individual integral is denoted as $M_{\cal G}$ 
and the sum of all $M_{\cal G}$ in Set~V is denoted as $M_{10}$.

%================================================================
\subsection{Construction of building blocks}
\label{subsec:buildingblocks}

The conversion of momentum integral into that of Feynman parameters
involves inversion of a large matrix,
which is performed by 
MAPLE.
This enables us to obtain explicit forms of
$A_i, B_{ij}, C_{ij}, U$ 
as homogeneous functions of $z_1, z_2, \ldots, z_9; z_a, z_b, \ldots, z_e$.
$V$ has a form given in Eq.~(\ref{def:V}) which is
common to all diagrams of Set~V.

%================================================================
\subsection{Construction of UV divergence subtraction terms}
\label{subsec:UVsubtractionterms}

The renormalization of UV divergence is carried out by a subtractive method.
A UV divergence of a diagram of Set~V arises 
from a subdiagram $\mathcal{S}$, which is of vertex
type or self-energy type.
The Set~V has no subdiagrams of
vacuum-polarization type or light-by-light-scattering type. 

Suppose $M_\mathcal{G}$
diverges when all loop momenta of a subdiagram $\mathcal{S}$
consisting of $N_{\mathcal{S}}$ lines and $n_{\mathcal{S}}$ closed loops
go to infinity.
In the Feynman-parametric formulation, this corresponds to the vanishing of
the denominator $U$ when all $z_i \in \mathcal{S}$ vanish simultaneously.
To find a criterion for a UV divergence from $\mathcal{S}$,
consider the part of the integration domain where $z_i$ satisfies
$\sum_{i \in \mathcal{S}} z_i \leq \epsilon$.  
In the limit $\epsilon \rightarrow 0$ one finds
\begin{eqnarray}
V &=&  {\cal O}(1),~~~~U = {\cal O}(\epsilon^{n_\mathcal{S}}),\nonumber \\
B_{ij} &=& {\cal O}(\epsilon^{n_\mathcal{S}-1} )~~~{\rm if}~~i,j \in \mathcal{S},\nonumber \\
B_{ij} &=& {\cal O}(\epsilon^{n_\mathcal{S}} )~~~{\rm otherwise}.
\label{UVlimit}
\end{eqnarray}
From this we can obtain a simple UV power-counting rule
for identifying UV divergent terms.
Based on this information we can construct an integral which has the same
UV divergence as $M_\mathcal{G}$ but has features suitable as 
the UV divergence counterterm.
The {\it K}-operation ${\mathbb K}_{\mathcal S}$
\cite{Cvitanovic:1974sv,Aoyama:2005kf} on $M_\mathcal{G}$  
that creates such a counterterm has properties summarized as follows:

(i)  The integral ${\mathbb K}_{\mathcal S}M_{\cal G}$ 
subtracts the UV divergence arising from the subdiagram 
$\mathcal{S}$ of $M_{\cal G}$ 
{\it point by point} in the same Feynman parametric space.

(ii)  By construction the subtraction term factorizes into 
pieces of magnetic moments and renormalization constants
of lower-order, which are known from lower-order calculation.

For a vertex-type subdiagram $\mathcal{S}$
the {\it K}-operation ${\mathbb K}_{\mathcal S}$ on $M_{\mathcal G}$
factorizes exactly into the product of lower-order quantities as
\begin{equation}
{\mathbb K}_{\mathcal S}  M_{\mathcal G}=L_{\mathcal S}^{UV} M_{\mathcal{G}/\mathcal{S}}.
\label{koponvertex}
\end{equation}
where ${\mathcal G}/{\mathcal S}$ is the reduced diagram obtained by shrinking ${\mathcal S}$ in ${\mathcal G}$ to a point,
and $L_{\mathcal S}^{UV}$ is the leading UV divergent part
of the vertex renormalization constant $L_{\mathcal S}$.

For a self-energy-type subdiagram $\mathcal{S}$,
connected to the rest of ${\mathcal G}$ by electron lines $i$ and $j$,
the {\it K}-operation ${\mathbb K}_{\mathcal S}$ on $M_{\mathcal G}$ gives two terms
of the form
\begin{equation}
{\mathbb K}_{\mathcal S} M_{\mathcal G}=dm_{\mathcal S}^{UV} 
M_{\mathcal{G}/\mathcal{S}(i^{*})} 
+ B_\mathcal{S}^{UV} M_{\mathcal{G}/[\mathcal{S},j]}.
\label{koponselfenergy}
\end{equation}
Here $M_{\mathcal{G}/\mathcal{S}(i^{*})}$  is the reduced diagram obtained by
shrinking ${\mathcal S}$ to a point and
$i^*$ indicates that the two-point mass vertex 
is inserted in the line $i$ of
the diagram $M_{\mathcal{G}/\mathcal{S}}$.
The second term comes from the
diagram obtained by shrinking both ${\mathcal S}$ and $j$ to a point.
This term, which is written as ${\mathcal G}/[{\mathcal S}, j]$,
where $[{\mathcal S}, j]$ denotes the sum of two sets,
can be transformed into a more convenient form 
by integration by part with respect to $z_i$.
$dm_{\mathcal S}^{UV}$ and  $B_{\mathcal S}^{UV}$ are the leading UV divergent parts of the mass renormalization constant
$dm_{\mathcal S}$ and wave-function renormalization constant $B_{\mathcal S}$.
See Ref.~\cite{Kinoshita:1990} for more details.

(iii)  The {\it K}-operation
generates only the leading UV-divergent parts
of renormalization constants.
Thus an additional {\it finite} renormalization 
(called residual renormalization)  is required to recover
the standard on-shell renormalization.

In general 
the subtracting integrand is derived from the original integrand
by applying several {\it K}-operations on the Zimmermann's forest of subdiagrams
\cite{Zimmermann:1969}.
Suppose ${\mathbb K}_\mathcal{S}$ is the {\it K}-operator associated with a subdiagram $\mathcal{S}$
of a diagram $\mathcal{G}$.  Then the UV-finite amplitude $M_\mathcal{G}^{\rm R}$ is
obtained from the unrenormalized amplitude $M_\mathcal{G}$ by the forest formula
of the form
\cite{Aoyama:2005kf}
\begin{equation}  
M_\mathcal{G}^{\rm R} = \sum_{f \in \mathfrak{F}(\mathcal{G})} \left [ \prod_{S_i \in f} \left ( - \mathbb{K}_{S_i}
\right ) \right ] M_\mathcal{G} ,
\label{eq:Kforest}
\end{equation}  
where the sum is taken over all forests $f$,
including an empty forest, of the diagram $\mathcal{G}$.
The order of operation in the product is arranged so that operations for 
the outer subdiagrams are applied first.

%================================================================
\subsection{Construction of IR divergence subtraction terms}
\label{subsec:IRsubtractionterms}

The IR divergence has its origin in 
the singularity caused by vanishing mass of virtual photon.
However, this is just a necessary condition, not a sufficient condition.
In order that this singularity causes actual IR divergence of the integral
it must be enhanced by vanishing of denominators of two or more electron 
propagators (called {\it enhancers}) due to kinematic constraints. 
Such a situation arises in the diagrams that have self-energy subdiagrams.
It is associated with the vanishing of $V$-function
of the denominators in Eq.~(\ref{M_G}) in the integration domain characterized 
by \cite{Cvitanovic:1974sv,Kinoshita:1990}
\begin{align}
  z_i &= {\cal O} (\delta)
  && \text{if $i$ is an electron line in $\mathcal{R}$, where $\mathcal{R}\equiv{\mathcal{G}}/{\mathcal{S}}$},  
\nonumber \\
  z_i &= {\cal O} (1)
  && \text{if $i$ is a photon line in $\mathcal{R}$},
\nonumber \\
  z_i &= {\cal O} (\epsilon),\ \epsilon \sim \delta^2,
  && \text{if $i \in \mathcal{S}$}.
\label{irlimit}
\end{align}

The origin of linear or higher power IR-divergence is easy to identify 
diagrammatically.
It is caused when
the diagram has two or more disconnected self-energy-like
subdiagrams and one of the self-energy-like subdiagrams
behaves as a self-mass when the photon momenta of the diagram
outside the self-energy-like subdiagram vanish and the electron
lines attached to it go on shell.

Our treatment of self-energy subdiagram
by means of the {\it K}-operation 
subtracts only the UV-divergent part of the self-mass
as is shown in Eq.~(\ref{koponselfenergy}).
The unsubtracted remainder of self-mass term is 
proportional to $M_{\mathcal{G}/\mathcal{S}(i^{*})}$,
which contains an IR divergence. 
(In the case of second-order self-mass this problem does not arise since the 
entire self-mass term is removed by the {\it K}-operation.)

In order to avoid this problem we developed a method,
called {\it R}-subtraction \cite{Aoyama:2007mn}, which 
removes the finite remnant of the self-mass term completely, 
wherever it arises in a diagram.
For a formal treatment, we introduce 
the {\it R}-subtraction operator $\mathbb{R}_\mathcal{S}$ 
\begin{equation}
  \mathbb{R}_\mathcal{S} {M}_\mathcal{G}
  =
  dm_{\mathcal{S}}^{\rm R}
  {M}_{\mathcal{G}/\mathcal{S}\,(i^\star)}^{\rm R}, 
\end{equation}
where ${dm}_{\mathcal{S}}^{\rm R}$ is the UV-finite part 
of the mass renormalization constant defined by 
\begin{equation}
  {dm}_{\mathcal{S}}^{\rm R}
  =
  {dm}_{\mathcal{S}}
  -
  {dm}_{\mathcal{S}}^{\rm UV}
  +
  \sum_{f} \prod_{\mathcal{S}^\prime\in f}
  \left(-\mathbb{K}_{\mathcal{S}^\prime}\right)
  \widetilde{dm}_{\mathcal{S}} ,
\label{dmR}
\end{equation}
and ${M}_{\mathcal{G}/\mathcal{S}\,(i^\star)}^{\rm R}$ 
is the UV-finite part extracted by means of the 
{\it K}-operation on the magnetic moment amplitude
of the residual diagram $\mathcal{G}/\mathcal{S}$ 
\begin{equation}
  {M}_{\mathcal{G}/\mathcal{S}\,(i^\star)}^{\rm R} 
  =
  \sum_{f} \prod_{\mathcal{S}^\prime \in f}
  \left( -\mathbb{K}_{\mathcal{S}^\prime} \right)
  {M}_{\mathcal{G}/\mathcal{S}\,(i^\star)},
\end{equation}
in which the leading UV-divergent part ${dm}_{\mathcal{S}}^{\rm UV}$ 
is entirely removed from the renormalization constant
${dm}_{\mathcal{S}}$
and the UV divergence in the remainder
$\widetilde{dm} \equiv {dm} - {dm}^{\rm UV}$
is subtracted away by applying the {\it K}-operation
associated with the forest $f$. 

The {\it R}-subtraction removes the power-law IR divergences
as well as logarithmic divergences related to self-mass. 
Another type of logarithmic IR divergence occurs, however,
 when the self-energy-like subdiagram ${\mathcal S}$ behaves 
as a lower-order magnetic moment and the residual factor 
${\mathcal G}/{\mathcal S}$ contains an IR singularity
analogous to the vertex renormalization constant of the diagram
${\mathcal G}/{\mathcal S}$.

By construction the resulting integral is factorizable into the product of
the magnetic moment $M_{\mathcal S}$ defined on the subset $\mathcal{S}$
and the UV-finite part 
${L}_{\mathcal{G}/\mathcal{S}}^{\rm R}$  
of the vertex renormalization constant 
${L}_{\mathcal{G}/\mathcal{S}}$ defined by  
\begin{equation}
  {L}_{\mathcal{G}/\mathcal{S}}^{\rm R}
  =
  {L}_{\mathcal{G}/\mathcal{S}}
  -
  {L}_{\mathcal{G}/\mathcal{S}}^{\rm UV}
  +
  \sum_{f} \prod_{\mathcal{S}^\prime\in f}
  \left(-\mathbb{K}_{\mathcal{S}^\prime}\right)
  \widetilde{L_{\phantom{G}}}_{\!\!\!\!\!\mathcal{G}/\mathcal{S}},
\label{LR}
\end{equation}
in which 
the leading UV-divergent part ${L}_{\mathcal{G}/\mathcal{S}}^{\rm UV}$ 
and the UV-divergent parts associated with the forests 
$\prod_{\mathcal{S}^\prime\in f} \left(-\mathbb{K}_{\mathcal{S}^\prime}\right) \widetilde{L_{\phantom{G}}}_{\!\!\!\!\!\mathcal{G}/\mathcal{S}}$
are subtracted away, where $\widetilde{L_{\phantom{G}}}\!\!\equiv L - L^{\rm UV}$. 

The {\it I}-subtraction operator $\mathbb{I}_\mathcal{S}$ acting 
on 
the unrenormalized amplitude $M_\mathcal{G}$ is defined by 
\begin{equation}
  \mathbb{I}_\mathcal{S} M_\mathcal{G}
  =
  {L}_{\mathcal{G}/\mathcal{S}}^{\rm R} M^{\rm R}_\mathcal{S}. 
\label{IoponMG}
\end{equation}

{\it
N.~B. The IR power counting rule identifies only IR-divergent part.
It does not specify how to handle the IR-finite part.
The {\it I}-subtraction operation 
defined by Eq.~(\ref{IoponMG})
handles the IR-finite terms in a manner different 
from that of the old subtraction method 
\cite{Cvitanovic:1974sv,Kinoshita:1990}.
Note also that the ``new'' {\it I}-subtraction operation applies only to
the self-energy-like subdiagram $\mathcal{S}$.
}

The whole set of IR subtraction terms can be obtained by 
the combination of 
{\it R}- and {\it I}-operations, 
both of which belong to {\it annotated forests}
\cite{Aoyama:2007bs}. 
An annotated forest is a set of self-energy-like subdiagrams, to 
each element of which the distinct operation of {\it I}-subtraction 
or {\it R}-subtraction is assigned. 
The IR-subtraction term associated with an annotated forest is 
constructed by successively applying operators $\mathbb{I}$ or 
$\mathbb{R}$, and takes the form 
\begin{equation}
  (-\mathbb{I}_{\mathcal{S}_i})\dots
  (-\mathbb{R}_{\mathcal{S}_j})\dots
   M_\mathcal{G}
\end{equation}
where the annotated forest consists of the subdiagrams 
$\mathcal{S}_i$, \dots and $\mathcal{S}_j$, \dots.

%================================================================
\subsection{Residual renormalization}
\label{subsec:Residualrenormalization}

The output of the steps {\bf A} through {\bf E}, 
which has been made UV-finite by {\it K}-operation and 
IR-finite by {\it R}- and {\it I}-operations, 
is not the standard-renormalized integral.
Thus an additional finite renormalization is  required
to obtain the standard result of on-shell-renormalization.

The sum of residual renormalization terms of all diagrams of Set~V 
is shown in Appendix~\ref{sec:appendixRes}.
The result can be written as the sum of terms,
all of which are free from UV- and IR-divergences:
\begin{align}
        A_1^{(10)} [\text{Set~V}]
	&= 
         \Delta M_{10} [\text{Set~V} ] 
\nonumber \\
        & + \Delta M_8 (-7 \DLB_2)
\nonumber \\
        & + \Delta M_6 \{ - 5 \DLB_4 + 20 (\DLB_2)^2 \}
\nonumber \\
	& + \Delta M_{4} \{ - 3 \DLB_6 
     +24 \DLB_4 \DLB_2 -28 (\DLB_2)^3 +2 \Delta L_{2^*} \Delta dm_4 \} 
\nonumber \\
       & + M_{2} \{ -\DLB_8 + 8 \DLB_6 \DLB_2 -28 \DLB_4 (\DLB_2)^2  
\nonumber \\
       & ~~~~~~~~     +  4 (\DLB_{4})^2 + 14 (\DLB_2)^4  + 2  \Delta dm_6  \Delta L_{2^*}  \}
\nonumber \\
       & + M_{2} \Delta dm_4 ( -16 \Delta L_{2^*} \DLB_2  + \Delta L_{4^*}  
                               -2 \Delta L_{2^*} \Delta dm_{2^*}  ),
	\label{a_Vr}
\end{align}
where 
$\Delta M_n$, $\DLB_n$, $\Delta dm_n$, $\Delta L_{n^*}$, and $\Delta dm_{2^*}$ 
are finite quantities of lower orders obtained 
in our calculation of lower order $a_e$.
(All these are quantities of {\it q}-type diagrams 
since subdiagrams of Set~V are all {\it q}-type.)
See Appendix~\ref{sec:appendixRes} for precise definitions.

%================================================================
\section{Numerical integration}
\label{sec:result}

We evaluate individual integrals by numerical integration 
using the iterative-adaptive Monte-Carlo routine
VEGAS \cite{Lepage:1977sw}.
A typical integrand consists of about 90,000 lines of FORTRAN codes
occupying more than 6 Megabytes.
The domain of integration 
is a 13-dimensional unit cube
($0 \leq x_i \leq 1$, $i=1,2,\dots,13$) onto which the hyperplane 
of 14 Feynman parameters (see Eq.~(\ref{def:dz})) is mapped.

In order to assure the credibility of results it is important
to understand the nature of error estimate generated by VEGAS.
An important feature of VEGAS is that its sampling points
of the integrand tend to accumulate after several iterations
 in the region where it gives the most contribution to the integral.
Errors encountered in our work arise primarily from the following three
features of our integrands:
\begin{enumerate}[(a)]
  \item Our integrands are singular on some boundary surface of the unit cube
because of vanishing of the denominators $U$ and/or $V$,
whether they are renormalized or not.
  \item Our renormalization is performed numerically on a computer, which means
mutual cancellation of $\infty$'s at every singular point in the 
domain of integration.
  \item The sheer size of the integrands makes it difficult to accumulate
a sufficient amount of sampling data with limited computing power available.
\end{enumerate}

\subsection{Steep landscape of integrands and stretching}

At first sight, the feature (a) seems to indicate that it is hopeless
to obtain a reliable result for such a singular integrand.
However, the measure of immediate neighborhood of singularity 
is small enough so
that the integral itself is well defined and convergent because 
of renormalization.

Nevertheless, steep landscape of the integrand may be a cause 
for concern since the grid adjustment by VEGAS might not reach
the optimal stage as rapidly as one would wish.
This problem may be alleviated, however, by ``stretching'' the integration
variables.
(See Sec.~6.3 of Ref.~\cite{Kinoshita:1990}.)

Suppose VEGAS finds after several iterations that sampling points
are highly concentrated at one end of the integration domain,
say $x_i = 0$, where $x_i$ is one of the axes of the hypercube.
In such a case, if one maps $x_i$ into $x_i^{\prime}$ as
\begin{equation}
x_i = {x_i^{\prime}}^{a_i} ,
\label{stretch0}
\end{equation}
where $a_i$ is some real number greater than 1,
the neighborhood of $x_i = 0$ is stretched out
and random samplings in $x_i^{\prime}$ give more attention to
the region near $x_i = 0$ from the beginning of iteration.
Also the Jacobian 
$a_i {x_i^{\prime}}^{a_i - 1}$ 
of the transformation (\ref{stretch0})
has the effect of reducing the peak of the integrand.
Similarly, the singularity at $x_i =  1$ can be weakened by the stretching
\begin{equation}
x_i =  1 - (1 - x_i^{\prime})^{b_i} ,
\label{stretch1}
\end{equation}
where $b_i$ is some real number greater than 1.

Stretching may be applied to all integration variables independently.
By an appropriate choice of parameters $a_1$, $a_2$,\dots and $b_1$, $b_2$,\dots
the convergence of iteration can be accelerated considerably.

Note that the stretching is not an attempt to simulate the integrand
itself. It is designed to reduce the size of peaks 
indicated by preceding iterations so that the 
sampling points become more evenly distributed
 throughout the transformed domain of integration.
It is easy to implement since it is applied to each axis independently.
Since there is no constraint on the choice of $a_i$ and $b_i$, except 
that they must be  real numbers larger than 1, 
one can try various stretches and choose more efficient one. 
Since different stretches are nothing but
evaluation of the same integral with different distributions of sampling points,
they can also be used to check the consistency of calculation.

\subsection{Extended numerical precision}

Concerning the feature (b),
the integrals are made convergent by point-by-point
cancellation of divergences by carefully tailored counterterms
created by the intermediate renormalization procedure.
All this would pose no problem if each step of computation were carried
out with infinite precision.
In practice, however, we have to perform calculations in finite precision.
The intended cancellation may fail occasionally because cancelling
terms have only finite number of significant digits, and their difference,
which is supposed to vanish at the singular point,
might be dominated near the singularity by roundoff errors,
causing uncontrolled fluctuation.
This problem can be reduced to a manageable level by adopting
higher precision arithmetics which will reduce the size of
dangerous integration volume, although it
slows down the computation severely.

In our calculation, some of the diagrams are evaluated 
in the double-double (pseudo-quadruple) precision arithmetic
using the library written by one of us 
which is the arrayed version of the algorithm presented in 
qd library \cite{Hida}.
For the diagram $X008$ that exhibits even more severe digit-deficiency, 
the most singular part of the integral is evaluated with 
quadruple-double (pseudo-octuple) precision 
and the remaining part is evaluated with the double-double precision. 

\subsection{Intensive computation}

The feature (c),
huge size of our integrands, means that they require a large amount
of computing time in order to accumulate sufficient sampling statistics.
Indeed, this combined with the difficulty in accessing adequate
computing resources has been the main cause of delay in
obtaining high accuracy result thus far.

\subsection{Numerical integration process and the result}

All integrals are evaluated initially in double precision 
using $10^7$ sampling points per iteration, 
iterated 50 times, followed by $10^8$ points per iteration,
iterated 50 times.
This step is to confirm that our renormalization scheme
actually works and gives finite results.

The output of {\sc gencode}{\it N},
being a universal code, employes a generic mapping of the Feynman parameters 
(denoted as the default mapping), 
and is not optimized for the individual diagrams. 

The first thing we must do to improve the convergence of iteration 
is to note that 
diagrams containing $n_s$ subdiagrams of self-energy type require
only $(13-n_s)$ independent integration variables.
The reduction of integration variables helps improve
the convergence of VEGAS iterations.
We shall call the class of these diagrams $X\!B$. 
It consists of 236 diagrams.
The remainder, which consists of 153 diagrams without self-energy
subdiagram, will be called $X\!L$.

Another improvement takes account of the fact that
the iteration of VEGAS converges better 
if singular behavior of the integrand is confined to
one axis.
For instance we may choose the largest sum of Feynman parameters 
that vanishes
at the singularity of the integrand as one to be mapped
onto an integration variable.
See discussion around Eq.~(\ref{UVlimit}).
This is not always possible for our integrands
which may have multiple sets of singular axes,
but it still helps.

After these adjustments are made,
each integral is evaluated in double precision arithmetic 
with $10^9$ sampling points, which takes 
1 to 3 hours on 32 cores of RICC (RIKEN Integrated Clusters of Clusters).
Evaluation in 
double-double (pseudo-quadruple) precision is about 60 times slower.
Some large runs in 
double-double precision with $10^9$ sampling points per iteration,
iterated 80 times, took about 65 days on the 128 cores of RICC.

Thus far $X\!L$-integrals were evaluated in two ways:
\begin{itemize}
\item[1)] Primary runs with the default mapping in double precision arithmetic. [$X\!L1$]
\item[2)] Second runs with the adjusted mapping in double precision arithmetic. [$X\!L2$]
\end{itemize}
$X\!B$-integrals were evaluated in three ways:
\begin{itemize}
\item[1)] Primary runs with the default mapping in double precision arithmetic. [$X\!B1$]
\item[2a)] Second run with the adjusted mapping in double-double precision arithmetic. There are 162 integrals. [$X\!B2$a]
\item[2b)] Remaining 74 integrals evaluated after the preliminary result was  published. [$X\!B2$b]
\item[3)] Third run in double precision for 176 of 236 integrals,
and double-double precision for the remaining 60 integrals. [$X\!B3$]
\end{itemize}

By early 2012 we managed to reduce the uncertainties of 
all individual integrals to less than 0.05.
The value of $A_1^{(10)}[\text{Set~V}]$
was obtained by combining $X\!L$1, $X\!B$1, and $X\!B$2a.
The combined uncertainty of $A_1^{(10)}[\text{Set~V}]$
was about 0.57.  
This was reported as a very preliminary value: 
\cite{ae10:PRL}
\begin{equation}
A_1^{(10)} [\text{Old Set~V}] = 10.092~(570).
\label{oldset5result}
\end{equation}

Since the preliminary result was published, 
we have reevaluated all tenth-order integrals for various choices of mapping.
The new result consists of $X\!L$2, $X\!B$2a, $X\!B$2b, and $X\!B$3, 
excluding $X\!L$1 and $X\!B$1.
They are summarized in Table~\ref{table:X001}. 
Auxiliary quantities required for the residual renormalization are
listed in Table~\ref{Table:residual_const}.
Combining them all we obtain
\begin{equation}
        A_1^{(10)} [\text{Set~V}] = 8.726~(336).
\label{set5result}
\end{equation}
The difference of new and old results 
is 1.366, which is twice as large as the combined uncertainty 0.662.
Another point to notice is that,
in spite of the far greater numbers of sampling points,
the uncertainty of (\ref{set5result}) is only 1.7 times smaller
than the uncertainty of the
very preliminary result (\ref{oldset5result}).

\subsection{Remarks}

In order to understand the possible cause of these results 
it is necessary to examine the behavior of individual integrals. 
VEGAS subdivides the integration domain into {\it grid}, from which 
sampling points of integrand are taken. 
The grid is adjusted adaptively based 
on the results of previous iterations so that the importance sampling 
is achieved. 
If the absolute value of the integrand has a peak, 
sampling points will accumulate in that neighborhood 
as the iteration progresses to accelerate the convergence. 

For the multivariate integration, the grid adjustment relies on 
the profile of the integral projected along each axis. 
It is monitored by the information that VEGAS provides after 
each iteration by printing out the values of the integrand at ten 
points along each axis integrated over the remaining variables, 
in addition to the value and error of the integral itself. 
However, some integrands may have several competing peaks. 
In such a case VEGAS might find initially only one peak, unaware 
of the presence of other peaks, 
if the number of sampling points is too small, 
and might be lead to unstable convergence, a misleading values, 
or an unreliable error estimate. 

It may occur that the grid adjustment does not work well when the 
peaks or singularities of the integrand are not localized along an axis 
but rather located, for instance, in the diagonal region over several axes.%
\footnote{%
A new version of VEGAS provided by P.~Lepage in 2013 overcomes
this known weakness of the original version of VEGAS \cite{Lepage:1977sw}.
The new VEGAS can be obtained from \url{https://github.com/gplepage/vegas}.
We have not used the new VEGAS algorithm, since it was released after 
we had carried out most of Set~V integration over several years.
}
In our calculation, the singular behavior of the integrand associated with 
the divergences lies at the boundaries of the integration domain. 
It should be desirable to choose integration variables so that 
the singularities are concentrated on one end of the axis, 
e.g., $x_i \to 0$ rather than the situation 
where they emerge, e.g., when the variables $x_i$ and $x_j$ 
go to zero simultaneously. 

We may note that 14 Feynman parameters $z_i$ of tenth order diagrams 
satisfying $\sum z_i = 1$ 
are mapped to the integration domain of a 13 dimensional unit cube. 
The choice of mapping is arbitrary, and thus the appropriate mapping 
should be applied that takes account of the above considerations, 
reflecting the substructure of the diagram. 
In general, the default mapping adopted in the output of {\sc gencode}{\it N}, 
being the universal code, is not optimal in this sense. 

The calculation runs $X\!L$1 and $X\!B$1 contributing 
to the preliminary result \cite{ae10:PRL} 
rely on the default mapping. 
Several integrals seems to suffer from some of the problems described above. 
By using the different mappings that are tailored for individual diagrams, 
especially for the diagrams of $X\!L$1 containing several second-order 
vertex subdiagrams, the convergence rates of the integral have been much 
improved and the reliable error estimates are obtained. 
This observation suggests that in the numerical integration of the 
tenth order diagrams, when the number of sampling points are not 
large enough, 
the inappropriate mapping would lead to some underestimate of the error 
because the evaluated integrands over the sampling points do not obey 
the Gaussian distribution. 
The uncertainty of Ref.~\cite{ae10:PRL} was thus not yet reliable and 
must be enlarged substantially. 
On the other hand, integrals contributing to the new result behave 
much better presumably because of the new mappings. The values and the 
error estimates are reliable also because of substantially 
increased sampling statistics. 

Now that the improved values of all diagrams of Set~V are obtained
we have a complete evaluation of 12,672 diagrams of tenth-order
\cite{Kinoshita:2005sm,Aoyama:2008gy,Aoyama:2008hz,Aoyama:2010yt,Aoyama:2010pk,Aoyama:2010zp,Aoyama:2011rm,Aoyama:2011zy,Aoyama:2011dy,Aoyama:2012fc}.
Taking (\ref{set5result}) into account we report
\begin{equation}
        A_1^{(10)} = 7.795~(336)
\label{allsetresult}
\end{equation}
as the new tenth-order term.
It is about 14 times more precise than
the crude estimate $ | A_1^{(10)} | < 4.6$ \cite{Mohr:2008fa}
and makes the overall {\it theoretical} uncertainty
about 7.5 times smaller than the current experimental uncertainty
\cite{Hanneke:2008tm,Hanneke:2010au}.

%================================================================
\section{Discussion}
\label{sec:discussion}

In view of the enormous size and complexity of the integrals of Set~V,
it is unlikely that the validity of our results can be tested 
by an independent method any time soon.
We are thus obliged to establish their validity 
to the best of our capability.

First of all we have to make sure that our formulation is analytically exact.
FORTRAN codes of all integrals of Set~V are created by the code-generating
algorithm {\sc gencode}{\it N}, which has been tested extensively
by applying it to the creation of lower-order
diagrams of {\it q}-type.
Recall that {\it N} of {\sc gencode}{\it N} 
represents the number of vertices of {\it q}-type
diagrams where virtual photons are attached.
For {\it N} = $10$ it generates a complete set of 
distinct irreducible diagrams of Set~V automatically.
Similarly, complete sets of distinct {\it q}-type diagrams
of sixth- or eighth-orders
are generated by {\sc gencode}{\it N} for {\it N} = $6$ or $8$.
Since these integrals
have been thoroughly tested by comparison with previous formulations, we may
expect that {\sc gencode}{\it N} works correctly for {\it N} = $10$, too. 
We have found, however, that 
the implementation of {\sc gencode}{\it N} for constructing 
some IR-subtraction terms of the diagrams {\rm X253} and {\rm X256} 
requires modifications according to the definition of the renormalization 
constants with two-point vertex insertion. 
Since these exceptions are minor, we have corrected them manually 
instead of rewriting {\sc gencode}{\it N} itself.
This problem and its correction is discussed in full detail 
in Appendix~\ref{sec:appendixX253}.
With this modification the FORTRAN codes of 389 integrals,
including residual renormalization terms,
give a fully renormalized and analytically exact formula of $A_1^{(10)}$
for the Set~V.

The only uncertainty of our results
thus arises from the numerical integration
by the Monte-Carlo integration routine VEGAS \cite{Lepage:1977sw}.
The reliability of VEGAS has been tested 
thoroughly by applying it to the evaluation of
thousands of complicated integrals of sixth-order and eighth-order.
In all these cases the error estimates 
obtained by VEGAS, based on random sampling of the integrand,
 are found to be very reliable,
provided that 
a sufficiently large number of sampling data is accumulated.
This is helped significantly by stretching.
Double-double precision arithmetic is used
whenever problem caused by digit deficiency is suspected.
Of course because of their gigantic size numerical integration 
is extremely time-consuming
and accumulation of sampling statistics is a slow process.  
Inspection of Table~\ref{table:X001}
suggests that some of the
integrals may benefit from more extensive samplings.
There is an ongoing effort to improve the sampling statistics.

\newpage

%----------------------------------------------------------------
\begingroup
 \squeezetable
 \setlength\LTleft{0pt}
 \setlength\LTright{0pt}
 \setlength\LTcapwidth{\textwidth}
 \begin{longtable*}{@{\hskip+2em}lccdc@{\hskip+2em}}
%================================================================
 \caption*{%
\small TABLE~\ref{table:X001}:
 VEGAS integration results of $X001$--$X389$ of the tenth-order Set V diagrams.
 The superscript $dd$ in the first column means that the integrand
 was evaluated 
 with the double-double (pseudo-quadruple) precision. 
 The superscript $qd$ on $ X008$ indicates that
 the most singular part of the integral $ X008$ is evaluated with 
 quadruple-double (pseudo-octuple) precision 
 and the remaining part is evaluated with the double-double precision.
 Other integrals without the superscript 
 were evaluated with the double precision.
 The second column shows the symbolic representation of the diagram. 
 The third column counts the number of subtraction terms. 
 The fourth column presents the value of the integral with error 
 in the last few digits in the parentheses. 
 The fifth column lists the total number of
 iterations evaluated with $10^9$ sampling points per iteration.
\label{table:X001}
}\\
\hline\hline
\multicolumn{1}{p{.15\textwidth}}{Diagram} &
\multicolumn{1}{p{.15\textwidth}}{Vertex repr.} &
\multicolumn{1}{p{.15\textwidth}}{No.~of subtr.~terms} &
\multicolumn{1}{p{.25\textwidth}}{\hskip+1em Value (Error) including $n_F$} &
\multicolumn{1}{p{.25\textwidth}}{No.~of iterations with $10^9$ \newline sampling points per iteration} \\
\hline
\endfirsthead
%----------------------------------------------------------------
 \caption*{%
\small TABLE~\ref{table:X001} (continued):
  VEGAS integration results of $X001$--$X389$ of the tenth-order 
  Set V diagrams.
}\\
\hline\hline
\multicolumn{1}{p{.15\textwidth}}{Diagram} &
\multicolumn{1}{p{.15\textwidth}}{Vertex repr.} &
\multicolumn{1}{p{.15\textwidth}}{No.~of subtr.~terms} &
\multicolumn{1}{p{.25\textwidth}}{\hskip+1em Value (Error) including $n_F$} &
\multicolumn{1}{p{.25\textwidth}}{No.~of iterations with $10^9$ \newline sampling points per iteration} \\
\hline
\endhead
%----------------------------------------------------------------
\hline
\endfoot
%----------------------------------------------------------------
\hline
\hline
\endlastfoot
%================================================================
$ X001^{  }$  & $ abacbdcede $  &    47  &  -0.1724~(  91) &     20\\
$ X002^{dd}$  & $ abaccddebe $  &    47  &  -5.9958~( 333) &     13\\
$ X003^{  }$  & $ abacdbcede $  &    19  &  -0.1057~(  52) &     10\\
$ X004^{dd}$  & $ abacdcdebe $  &    71  &   5.1027~( 339) &      9\\
$ X005^{  }$  & $ abacddbece $  &    43  &   1.1112~( 168) &     20\\
$ X006^{  }$  & $ abacddcebe $  &    59  &  -5.2908~( 245) &      9\\
$ X007^{  }$  & $ abbcadceed $  &    47  &  -3.4592~( 254) &     25\\
$ X008^{qd}$  & $ abbccddeea $  &    47  & -16.5070~( 289) &     11\\
$ X009^{  }$  & $ abbcdaceed $  &    19  &  -3.1069~(  71) &     24\\
$ X010^{dd}$  & $ abbcdcdeea $  &    83  &  11.2644~( 342) &    124\\
$ X011^{dd}$  & $ abbcddaeec $  &    43  &   6.0467~( 338) &     22\\
$ X012^{dd}$  & $ abbcddceea $  &    67  &  -9.3328~( 267) &     26\\
$ X013^{  }$  & $ abcabdecde $  &     7  &  -1.3710~(  31) &      2\\
$ X014^{  }$  & $ abcacdedbe $  &    31  &   0.8727~(  42) &     10\\
$ X015^{  }$  & $ abcadbecde $  &     2  &   2.1090~(   8) &      2\\
$ X016^{  }$  & $ abcadcedbe $  &     2  &  -0.9591~(   7) &      2\\
$ X017^{  }$  & $ abcaddebce $  &     6  &   0.5146~(  13) &     20\\
$ X018^{  }$  & $ abcaddecbe $  &     6  &   0.0309~(  13) &     20\\
$ X019^{  }$  & $ abcbadeced $  &    31  &   1.2965~(  48) &     10\\
$ X020^{dd}$  & $ abcbcdedea $  &   134  &  -8.1900~( 318) &     43\\
$ X021^{  }$  & $ abcbdaeced $  &    11  &  -0.2948~(  15) &     10\\
$ X022^{  }$  & $ abcbdcedea $  &    79  &   0.8892~( 226) &     22\\
$ X023^{  }$  & $ abcbddeaec $  &    27  &   0.4485~(  55) &     25\\
$ X024^{  }$  & $ abcbddecea $  &    75  &  -6.0902~( 246) &     23\\
$ X025^{  }$  & $ abccadeebd $  &    39  &  -0.7482~( 194) &     20\\
$ X026^{dd}$  & $ abccbdeeda $  &    95  &  -7.8258~( 277) &      8\\
$ X027^{  }$  & $ abccdaeebd $  &    15  &  -2.3260~(  54) &     13\\
$ X028^{dd}$  & $ abccdbeeda $  &    71  &   4.5663~( 342) &     49\\
$ X029^{dd}$  & $ abccddeeab $  &    35  &   6.9002~( 233) &      1\\
$ X030^{dd}$  & $ abccddeeba $  &    67  & -12.6225~( 342) &     34\\
$ X031^{  }$  & $ abcdaebcde $  &     2  &   2.3000~(  14) &      4\\
$ X032^{  }$  & $ abcdaecdbe $  &     2  &  -0.2414~(   6) &      2\\
$ X033^{  }$  & $ abcdaedbce $  &     2  &  -1.3806~(   7) &      2\\
$ X034^{  }$  & $ abcdaedcbe $  &     2  &   1.2585~(   9) &      4\\
$ X035^{  }$  & $ abcdbeaced $  &     2  &  -0.5899~(   3) &      2\\
$ X036^{  }$  & $ abcdbecdea $  &    11  &   0.2318~(  11) &     30\\
$ X037^{  }$  & $ abcdbedaec $  &     2  &  -0.7407~(   5) &      2\\
$ X038^{  }$  & $ abcdbedcea $  &    11  &  -0.2927~(  14) &     20\\
$ X039^{  }$  & $ abcdceaebd $  &    11  &   0.3292~(  12) &     10\\
$ X040^{  }$  & $ abcdcebeda $  &    47  &   1.3397~(  50) &     12\\
$ X041^{  }$  & $ abcdcedeab $  &    63  &   3.1076~(  94) &     25\\
$ X042^{  }$  & $ abcdcedeba $  &   119  &  -4.1353~( 192) &     20\\
$ X043^{  }$  & $ abcddeeabc $  &    15  &  -2.9620~(  29) &     21\\
$ X044^{dd}$  & $ abcddeebca $  &    59  &   4.4121~( 281) &      4\\
$ X045^{  }$  & $ abcddeecab $  &    43  &   3.4331~( 212) &     20\\
$ X046^{dd}$  & $ abcddeecba $  &    95  &  -7.7564~( 339) &     15\\
$ X047^{  }$  & $ abcdeabcde $  &     2  &  -4.4496~(  40) &      8\\
$ X048^{  }$  & $ abcdeacdbe $  &     2  &  -0.8061~(   8) &      2\\
$ X049^{  }$  & $ abcdeadbce $  &     2  &  -0.0278~(   7) &      2\\
$ X050^{  }$  & $ abcdeadcbe $  &     2  &  -1.2213~(   9) &      4\\
$ X051^{  }$  & $ abcdebaced $  &     2  &  -0.1776~(   6) &      2\\
$ X052^{  }$  & $ abcdebcdea $  &    11  &   1.0293~(  17) &     20\\
$ X053^{  }$  & $ abcdebdaec $  &     2  &   0.3699~(   4) &      2\\
$ X054^{  }$  & $ abcdebdcea $  &    11  &  -0.5174~(  11) &     20\\
$ X055^{  }$  & $ abcdecaebd $  &     2  &  -0.3673~(   4) &      2\\
$ X056^{  }$  & $ abcdecbeda $  &    11  &  -0.2650~(  27) &     20\\
$ X057^{  }$  & $ abcdecdeab $  &    23  &   2.7370~(  31) &     30\\
$ X058^{  }$  & $ abcdecdeba $  &    44  &  -5.2510~(  70) &     12\\
$ X059^{  }$  & $ abcdedeabc $  &    23  &   2.1866~(  28) &     30\\
$ X060^{  }$  & $ abcdedebca $  &    92  &  -3.2089~( 188) &     22\\
$ X061^{  }$  & $ abcdedecab $  &    68  &  -3.7724~( 137) &     20\\
$ X062^{  }$  & $ abcdedecba $  &   161  &   5.9174~( 262) &     26\\
$ X063^{  }$  & $ abcdeeabcd $  &     6  &   3.4295~(  14) &     20\\
$ X064^{  }$  & $ abcdeeacbd $  &     6  &  -0.2772~(   8) &     20\\
$ X065^{  }$  & $ abcdeebadc $  &     6  &   0.1551~(  13) &     20\\
$ X066^{  }$  & $ abcdeebcda $  &    26  &  -3.6145~(  45) &     21\\
$ X067^{  }$  & $ abcdeecdab $  &    50  &  -1.6761~(  85) &     25\\
$ X068^{  }$  & $ abcdeecdba $  &    98  &   2.7855~( 217) &     22\\
$ X069^{  }$  & $ abcdeedabc $  &    18  &  -1.2627~(  31) &     11\\
$ X070^{  }$  & $ abcdeedbca $  &    70  &   3.2149~( 144) &     20\\
$ X071^{  }$  & $ abcdeedcab $  &    54  &   3.7025~(  96) &     20\\
$ X072^{  }$  & $ abcdeedcba $  &   134  &  -5.5704~( 208) &     15\\
$ X073^{  }$  & $ abacbdceed $  &    47  &   3.4114~( 254) &     24\\
$ X074^{  }$  & $ abacbddece $  &    47  &   4.4104~( 251) &     49\\
$ X075^{dd}$  & $ abacbddeec $  &    47  &  -8.1138~( 340) &     33\\
$ X076^{  }$  & $ abacbdecde $  &    19  &  -5.3405~(  74) &     26\\
$ X077^{  }$  & $ abacbdeced $  &    39  &   3.5459~(  86) &     56\\
$ X078^{  }$  & $ abacbdedce $  &    39  &   1.1666~(  80) &     56\\
$ X079^{  }$  & $ abacbdedec $  &    71  &   5.3956~( 305) &     41\\
$ X080^{  }$  & $ abacbdeecd $  &    43  &   0.4597~( 257) &     28\\
$ X081^{  }$  & $ abacbdeedc $  &    59  &  -5.6566~( 248) &     26\\
$ X082^{dd}$  & $ abaccdbeed $  &    47  &  -8.5156~( 348) &     92\\
$ X083^{dd}$  & $ abaccddeeb $  &    47  &  18.7464~( 346) &    117\\
$ X084^{  }$  & $ abaccdebde $  &    19  &   8.9888~( 129) &     20\\
$ X085^{  }$  & $ abaccdebed $  &    39  &  -2.2833~( 197) &     20\\
$ X086^{  }$  & $ abaccdedbe $  &    39  &   0.5180~( 223) &     20\\
$ X087^{dd}$  & $ abaccdedeb $  &    77  & -16.5849~( 349) &    160\\
$ X088^{dd}$  & $ abaccdeebd $  &    43  &  -5.2606~( 340) &     58\\
$ X089^{dd}$  & $ abaccdeedb $  &    63  &  12.6789~( 341) &     59\\
$ X090^{  }$  & $ abacdbceed $  &    19  &   1.5206~( 130) &     20\\
$ X091^{  }$  & $ abacdbdece $  &    39  &  -1.6355~(  97) &     56\\
$ X092^{  }$  & $ abacdbdeec $  &    39  &   2.1303~( 218) &     15\\
$ X093^{  }$  & $ abacdbecde $  &     7  &  -1.7594~(  42) &     10\\
$ X094^{  }$  & $ abacdbeced $  &    15  &  -1.0419~(  66) &     10\\
$ X095^{  }$  & $ abacdbedce $  &     7  &   0.5838~(  35) &      6\\
$ X096^{  }$  & $ abacdbedec $  &    31  &   1.3458~(  73) &     10\\
$ X097^{  }$  & $ abacdbeecd $  &    17  &   5.0319~(  89) &     24\\
$ X098^{  }$  & $ abacdbeedc $  &    33  &  -1.9806~( 183) &     20\\
$ X099^{  }$  & $ abacdcbeed $  &    39  &   3.0771~( 187) &     20\\
$ X100^{dd}$  & $ abacdcdeeb $  &    77  & -15.2919~( 331) &    244\\
$ X101^{  }$  & $ abacdcebde $  &    15  &  -0.2462~(  64) &     12\\
$ X102^{  }$  & $ abacdcebed $  &    31  &  -1.2883~(  75) &     26\\
$ X103^{  }$  & $ abacdcedbe $  &    31  &   0.9424~(  74) &     10\\
$ X104^{  }$  & $ abacdcedeb $  &    79  &   6.4131~( 298) &     42\\
$ X105^{  }$  & $ abacdceebd $  &    35  &   3.0503~( 215) &     21\\
$ X106^{  }$  & $ abacdceedb $  &    71  & -11.5662~( 344) &     48\\
$ X107^{dd}$  & $ abacddbeec $  &    43  &  -4.6573~( 345) &     77\\
$ X108^{dd}$  & $ abacddceeb $  &    63  &  12.9775~( 341) &     58\\
$ X109^{  }$  & $ abacddebce $  &    17  &  -0.0860~(  85) &     25\\
$ X110^{  }$  & $ abacddebec $  &    35  &   1.9248~( 204) &     20\\
$ X111^{  }$  & $ abacddecbe $  &    33  &   3.3578~( 132) &     24\\
$ X112^{  }$  & $ abacddeceb $  &    71  & -11.8998~( 332) &     53\\
$ X113^{dd}$  & $ abacddeebc $  &    39  &  -4.3847~( 322) &     16\\
$ X114^{dd}$  & $ abacddeecb $  &    63  &  11.0641~( 343) &     54\\
$ X115^{  }$  & $ abacdebcde $  &     7  &  -0.5974~(  52) &     12\\
$ X116^{  }$  & $ abacdebced $  &     7  &   1.8362~(  28) &     10\\
$ X117^{  }$  & $ abacdebdce $  &     7  &   0.3292~(  27) &     10\\
$ X118^{  }$  & $ abacdebdec $  &    15  &  -3.2721~(  55) &     10\\
$ X119^{  }$  & $ abacdebecd $  &    15  &  -0.0751~(  53) &     10\\
$ X120^{  }$  & $ abacdebedc $  &    31  &   1.8769~(  72) &     10\\
$ X121^{  }$  & $ abacdecbde $  &     7  &  -0.8549~(  43) &      6\\
$ X122^{  }$  & $ abacdecbed $  &     7  &  -0.7337~(  42) &      6\\
$ X123^{  }$  & $ abacdecdbe $  &    15  &  -3.3559~(  67) &     12\\
$ X124^{  }$  & $ abacdecdeb $  &    29  &  11.5746~( 106) &     26\\
$ X125^{  }$  & $ abacdecebd $  &    31  &   0.8677~(  64) &     10\\
$ X126^{  }$  & $ abacdecedb $  &    59  &  -1.5696~( 162) &     26\\
$ X127^{  }$  & $ abacdedbce $  &    15  &   1.1412~(  46) &     10\\
$ X128^{  }$  & $ abacdedbec $  &    31  &   0.6493~(  59) &     10\\
$ X129^{  }$  & $ abacdedcbe $  &    31  &   1.4833~(  70) &     10\\
$ X130^{  }$  & $ abacdedceb $  &    59  &  -1.5696~( 180) &     20\\
$ X131^{  }$  & $ abacdedebc $  &    59  &   3.1060~( 287) &     33\\
$ X132^{dd}$  & $ abacdedecb $  &   101  &  -8.8300~( 337) &     43\\
$ X133^{  }$  & $ abacdeebcd $  &    17  &   2.7263~(  88) &     24\\
$ X134^{  }$  & $ abacdeebdc $  &    33  &  -0.6712~( 123) &     23\\
$ X135^{  }$  & $ abacdeecbd $  &    33  &   0.9256~( 153) &     22\\
$ X136^{  }$  & $ abacdeecdb $  &    65  &  -7.5256~( 305) &     46\\
$ X137^{  }$  & $ abacdeedbc $  &    45  &  -2.3541~( 233) &     23\\
$ X138^{  }$  & $ abacdeedcb $  &    85  &  10.1610~( 284) &     38\\
$ X139^{dd}$  & $ abbcaddeec $  &    47  &  14.8650~( 348) &    104\\
$ X140^{  }$  & $ abbcadeced $  &    39  &  -2.7901~( 206) &     21\\
$ X141^{dd}$  & $ abbcadedec $  &    74  & -12.5567~( 350) &    261\\
$ X142^{dd}$  & $ abbcadeecd $  &    43  &  -1.5767~( 341) &     66\\
$ X143^{dd}$  & $ abbcadeedc $  &    61  &  10.3225~( 341) &     58\\
$ X144^{dd}$  & $ abbccdedea $  &    83  &  23.7239~( 368) &    230\\
$ X145^{dd}$  & $ abbccdeeda $  &    67  & -18.6212~( 349) &    115\\
$ X146^{dd}$  & $ abbcdadeec $  &    39  &  -2.2990~( 335) &     25\\
$ X147^{  }$  & $ abbcdaeced $  &    15  &   1.1243~(  55) &     20\\
$ X148^{  }$  & $ abbcdaedec $  &    31  &  -1.4150~(  76) &     21\\
$ X149^{  }$  & $ abbcdaeecd $  &    17  &  -8.3898~( 139) &     19\\
$ X150^{dd}$  & $ abbcdaeedc $  &    33  &   2.8758~( 260) &      2\\
$ X151^{dd}$  & $ abbcdcedea $  &    87  & -10.9362~( 344) &     68\\
$ X152^{dd}$  & $ abbcdceeda $  &    77  &  14.6793~( 345) &    113\\
$ X153^{dd}$  & $ abbcddecea $  &    77  &  14.8936~( 343) &     80\\
$ X154^{dd}$  & $ abbcddeeca $  &    67  & -20.6285~( 342) &     90\\
$ X155^{  }$  & $ abbcdeadec $  &    15  &   5.0341~(  46) &     20\\
$ X156^{  }$  & $ abbcdeaedc $  &    31  &  -0.8277~(  69) &     14\\
$ X157^{  }$  & $ abbcdecdea $  &    32  & -11.8490~( 252) &     18\\
$ X158^{dd}$  & $ abbcdeceda $  &    65  &   0.4607~( 329) &      6\\
$ X159^{  }$  & $ abbcdedcea $  &    65  &   0.4435~( 351) &     27\\
$ X160^{dd}$  & $ abbcdedeca $  &   116  &  14.0724~( 349) &    176\\
$ X161^{dd}$  & $ abbcdeecda $  &    71  &   7.8073~( 342) &     68\\
$ X162^{dd}$  & $ abbcdeedca $  &    95  & -12.8293~( 339) &     43\\
$ X163^{  }$  & $ abcabdceed $  &    19  &   6.8168~( 202) &     21\\
$ X164^{dd}$  & $ abcabddeec $  &    19  & -12.8880~( 208) &      3\\
$ X165^{  }$  & $ abcabdeced $  &    15  &  -2.1661~(  76) &     10\\
$ X166^{  }$  & $ abcabdedce $  &    15  &  -2.3080~(  70) &     10\\
$ X167^{  }$  & $ abcabdedec $  &    29  &  12.1361~( 150) &     20\\
$ X168^{  }$  & $ abcabdeecd $  &    17  &   3.4447~( 120) &     24\\
$ X169^{  }$  & $ abcabdeedc $  &    25  &  -6.9379~( 108) &     20\\
$ X170^{  }$  & $ abcacdbeed $  &    39  &   0.2635~( 288) &     36\\
$ X171^{dd}$  & $ abcacddeeb $  &    39  &  -2.5229~( 313) &      7\\
$ X172^{  }$  & $ abcacdebed $  &    31  &   1.5601~(  76) &     26\\
$ X173^{  }$  & $ abcacdedeb $  &    59  &   0.0193~( 298) &     48\\
$ X174^{  }$  & $ abcacdeebd $  &    35  &   1.7158~( 191) &     25\\
$ X175^{  }$  & $ abcacdeedb $  &    51  &  -1.8253~( 175) &     19\\
$ X176^{  }$  & $ abcadbceed $  &     7  &   0.7450~(  35) &     20\\
$ X177^{  }$  & $ abcadbdeec $  &    15  &   0.0079~(  81) &     21\\
$ X178^{  }$  & $ abcadbeced $  &     5  &   0.7159~(  28) &      2\\
$ X179^{  }$  & $ abcadbedce $  &     2  &  -0.4377~(   8) &      4\\
$ X180^{  }$  & $ abcadbedec $  &    11  &   0.0284~(  25) &      4\\
$ X181^{  }$  & $ abcadbeecd $  &     6  &  -4.4372~(  28) &     30\\
$ X182^{  }$  & $ abcadbeedc $  &    12  &   1.2822~(  43) &     20\\
$ X183^{  }$  & $ abcadcbeed $  &     7  &  -0.0791~(  29) &     20\\
$ X184^{  }$  & $ abcadcdeeb $  &    31  &   0.1973~( 134) &     25\\
$ X185^{  }$  & $ abcadcebed $  &     5  &  -0.1269~(  16) &     10\\
$ X186^{  }$  & $ abcadcedeb $  &    23  &   1.1883~(  21) &     10\\
$ X187^{  }$  & $ abcadceebd $  &     6  &   1.2699~(  27) &     20\\
$ X188^{  }$  & $ abcadceedb $  &    24  &   1.7966~(  36) &     11\\
$ X189^{  }$  & $ abcaddbeec $  &    17  &  -3.7500~( 105) &     20\\
$ X190^{  }$  & $ abcaddceeb $  &    33  &  -2.4966~( 217) &     20\\
$ X191^{  }$  & $ abcaddebec $  &    13  &   0.1892~(  62) &     11\\
$ X192^{  }$  & $ abcaddeceb $  &    25  &   2.3868~(  91) &     24\\
$ X193^{  }$  & $ abcaddeebc $  &    15  &  -4.2570~(  84) &     19\\
$ X194^{  }$  & $ abcaddeecb $  &    27  &  -0.6785~( 102) &     25\\
$ X195^{  }$  & $ abcadebcde $  &     2  &  -1.0708~(  19) &     10\\
$ X196^{  }$  & $ abcadebced $  &     2  &  -2.0432~(  20) &      6\\
$ X197^{  }$  & $ abcadebdce $  &     2  &  -0.3848~(   8) &      2\\
$ X198^{  }$  & $ abcadebdec $  &     5  &  -2.3533~(  26) &      2\\
$ X199^{  }$  & $ abcadebecd $  &     5  &   1.0636~(  26) &      2\\
$ X200^{  }$  & $ abcadebedc $  &    11  &   0.0266~(  26) &      4\\
$ X201^{  }$  & $ abcadecbde $  &     2  &  -0.4897~(  18) &      6\\
$ X202^{  }$  & $ abcadecbed $  &     2  &   1.9313~(  17) &      6\\
$ X203^{  }$  & $ abcadecdbe $  &     2  &   0.9061~(  10) &      4\\
$ X204^{  }$  & $ abcadecdeb $  &    11  &  -1.9485~(  26) &      2\\
$ X205^{  }$  & $ abcadecebd $  &     5  &  -0.9039~(  13) &     10\\
$ X206^{  }$  & $ abcadecedb $  &    23  &   1.6836~(  23) &     10\\
$ X207^{  }$  & $ abcadedbce $  &     5  &   0.2908~(  23) &      2\\
$ X208^{  }$  & $ abcadedbec $  &    11  &   0.5283~(  28) &      2\\
$ X209^{  }$  & $ abcadedcbe $  &     5  &   0.1496~(  19) &      2\\
$ X210^{  }$  & $ abcadedceb $  &    23  &   0.7803~(  19) &     10\\
$ X211^{  }$  & $ abcadedebc $  &    23  &   5.1339~(  90) &     12\\
$ X212^{  }$  & $ abcadedecb $  &    41  &  -0.4617~( 138) &     25\\
$ X213^{  }$  & $ abcadeebcd $  &     6  &  -2.4516~(  29) &     20\\
$ X214^{  }$  & $ abcadeebdc $  &    12  &   0.6801~(  39) &     20\\
$ X215^{  }$  & $ abcadeecbd $  &     6  &   0.0724~(  24) &     20\\
$ X216^{  }$  & $ abcadeecdb $  &    24  &  -1.3029~(  42) &     12\\
$ X217^{  }$  & $ abcadeedbc $  &    18  &  -2.2261~(  71) &     15\\
$ X218^{  }$  & $ abcadeedcb $  &    30  &  -1.6396~(  84) &     25\\
$ X219^{dd}$  & $ abcbaddeec $  &    39  &   1.3579~( 311) &      5\\
$ X220^{  }$  & $ abcbadedec $  &    59  &  -2.5734~( 222) &     27\\
$ X221^{  }$  & $ abcbadeecd $  &    35  &   0.6650~( 161) &     20\\
$ X222^{  }$  & $ abcbadeedc $  &    51  &   0.8293~( 178) &     20\\
$ X223^{dd}$  & $ abcbcdeeda $  &   116  &  17.5168~( 349) &    128\\
$ X224^{  }$  & $ abcbdadeec $  &    31  &   2.4729~( 110) &     20\\
$ X225^{  }$  & $ abcbdaedec $  &    23  &   0.3434~(  39) &     10\\
$ X226^{  }$  & $ abcbdaeecd $  &    13  &   1.0443~(  58) &     11\\
$ X227^{  }$  & $ abcbdaeedc $  &    25  &   0.5835~(  97) &     21\\
$ X228^{  }$  & $ abcbdceeda $  &    75  &  -6.8113~( 333) &     52\\
$ X229^{dd}$  & $ abcbddaeec $  &    35  &  -1.9843~( 323) &     11\\
$ X230^{dd}$  & $ abcbddeeca $  &    71  &  15.6844~( 350) &    115\\
$ X231^{  }$  & $ abcbdeadec $  &    11  &  -0.7737~(  28) &     10\\
$ X232^{  }$  & $ abcbdeaedc $  &    23  &   0.4608~(  38) &     10\\
$ X233^{  }$  & $ abcbdecdea $  &    31  &   8.6698~( 116) &     25\\
$ X234^{  }$  & $ abcbdeceda $  &    63  &  -2.5793~( 179) &     21\\
$ X235^{  }$  & $ abcbdedaec $  &    23  &   0.7486~(  35) &     10\\
$ X236^{  }$  & $ abcbdedcea $  &    63  &   2.0560~( 180) &     20\\
$ X237^{  }$  & $ abcbdedeca $  &   113  & -12.9913~( 363) &    154\\
$ X238^{  }$  & $ abcbdeeadc $  &    25  &   1.2747~(  45) &     21\\
$ X239^{  }$  & $ abcbdeecda $  &    69  &  -2.8075~( 345) &     49\\
$ X240^{  }$  & $ abcbdeedca $  &    93  &  10.9428~( 298) &     55\\
$ X241^{dd}$  & $ abccaddeeb $  &    43  &  13.8142~( 357) &    134\\
$ X242^{  }$  & $ abccadedeb $  &    68  & -10.4867~( 377) &    183\\
$ X243^{dd}$  & $ abccadeedb $  &    57  &   3.8891~( 336) &     44\\
$ X244^{dd}$  & $ abccdadeeb $  &    35  &  -3.3041~( 334) &     10\\
$ X245^{  }$  & $ abccdaedeb $  &    27  &   0.0658~(  83) &     12\\
$ X246^{  }$  & $ abccdaeedb $  &    29  &  -0.3959~( 174) &     20\\
$ X247^{dd}$  & $ abccddaeeb $  &    39  &  15.9539~( 344) &     43\\
$ X248^{dd}$  & $ abccddeaeb $  &    31  &  -1.9165~( 278) &      2\\
$ X249^{  }$  & $ abccdeadeb $  &    13  &   4.0116~(  46) &     20\\
$ X250^{  }$  & $ abccdeaedb $  &    27  &  -1.0558~(  68) &     24\\
$ X251^{  }$  & $ abccdedaeb $  &    27  &  -1.3906~(  76) &     12\\
$ X252^{dd}$  & $ abccdedeab $  &    56  & -10.9091~( 343) &     31\\
$ X253^{dd}$  & $ abccdedeba $  &   113  &  17.8437~( 352) &    221\\
$ X254^{  }$  & $ abccdeeadb $  &    29  &   2.2265~( 175) &     20\\
$ X255^{dd}$  & $ abccdeedab $  &    43  &   8.1598~( 340) &      6\\
$ X256^{dd}$  & $ abccdeedba $  &    93  & -14.0405~( 342) &     81\\
$ X257^{  }$  & $ abcdabceed $  &     7  &   5.7475~(  51) &     11\\
$ X258^{  }$  & $ abcdabdeec $  &     7  &  -0.5254~(  39) &     20\\
$ X259^{  }$  & $ abcdabeced $  &     5  &   0.0053~(  27) &     10\\
$ X260^{  }$  & $ abcdabedec $  &     5  &  -0.3958~(  20) &      2\\
$ X261^{  }$  & $ abcdabeecd $  &     6  &   6.4046~(  30) &     20\\
$ X262^{  }$  & $ abcdabeedc $  &     6  &  -2.2854~(  24) &     20\\
$ X263^{  }$  & $ abcdacbeed $  &     7  &  -2.8330~(  35) &     20\\
$ X264^{  }$  & $ abcdacdeeb $  &    15  &   4.8826~(  64) &     12\\
$ X265^{  }$  & $ abcdacebed $  &     5  &  -0.6756~(  20) &      2\\
$ X266^{  }$  & $ abcdacedeb $  &    11  &   0.1206~(  23) &     10\\
$ X267^{  }$  & $ abcdaceebd $  &     6  &  -0.6608~(  19) &     20\\
$ X268^{  }$  & $ abcdaceedb $  &    12  &   0.1185~(  31) &     20\\
$ X269^{  }$  & $ abcdadbeec $  &    15  &  -0.7190~(  56) &     12\\
$ X270^{  }$  & $ abcdadceeb $  &    31  &  -1.6881~(  97) &     25\\
$ X271^{  }$  & $ abcdadebec $  &    11  &   0.2492~(  23) &     10\\
$ X272^{  }$  & $ abcdadeceb $  &    23  &  -0.7285~(  32) &     10\\
$ X273^{  }$  & $ abcdadeebc $  &    13  &  -2.0474~(  45) &     11\\
$ X274^{  }$  & $ abcdadeecb $  &    25  &   0.8675~(  72) &     24\\
$ X275^{  }$  & $ abcdaebced $  &     2  &  -0.7496~(  12) &     10\\
$ X276^{  }$  & $ abcdaebdce $  &     2  &  -0.5547~(  10) &      4\\
$ X277^{  }$  & $ abcdaebdec $  &     2  &   2.7936~(  10) &      4\\
$ X278^{  }$  & $ abcdaebecd $  &     5  &  -0.1577~(  23) &     10\\
$ X279^{  }$  & $ abcdaebedc $  &     5  &   0.8399~(  15) &      2\\
$ X280^{  }$  & $ abcdaecbed $  &     2  &  -1.0127~(   8) &     10\\
$ X281^{  }$  & $ abcdaecdeb $  &     5  &  -1.3732~(  25) &      2\\
$ X282^{  }$  & $ abcdaecebd $  &     5  &   0.4907~(  18) &      2\\
$ X283^{  }$  & $ abcdaecedb $  &    11  &  -0.0427~(  23) &      2\\
$ X284^{  }$  & $ abcdaedbec $  &     2  &  -0.2670~(   9) &      2\\
$ X285^{  }$  & $ abcdaedceb $  &     5  &   0.0271~(  16) &      2\\
$ X286^{  }$  & $ abcdaedebc $  &    11  &   0.8014~(  21) &      2\\
$ X287^{  }$  & $ abcdaedecb $  &    23  &   0.2013~(  19) &     10\\
$ X288^{  }$  & $ abcdaeebcd $  &     6  &   4.2112~(  28) &     20\\
$ X289^{  }$  & $ abcdaeebdc $  &     6  &  -1.5651~(  19) &     20\\
$ X290^{  }$  & $ abcdaeecbd $  &     6  &  -3.7763~(  23) &     20\\
$ X291^{  }$  & $ abcdaeecdb $  &    12  &   1.5957~(  32) &     20\\
$ X292^{  }$  & $ abcdaeedbc $  &    12  &   0.9114~(  36) &     20\\
$ X293^{  }$  & $ abcdaeedcb $  &    24  &  -1.2653~(  41) &     11\\
$ X294^{  }$  & $ abcdbaceed $  &     7  &  -3.3891~(  25) &     20\\
$ X295^{  }$  & $ abcdbadeec $  &     7  &   1.7883~(  26) &     20\\
$ X296^{  }$  & $ abcdbaeced $  &     5  &   0.5511~(  13) &     10\\
$ X297^{  }$  & $ abcdbaedec $  &     5  &  -0.4696~(  16) &     10\\
$ X298^{  }$  & $ abcdbaeecd $  &     6  &  -1.9142~(  28) &     20\\
$ X299^{  }$  & $ abcdbaeedc $  &     6  &  -0.2907~(  22) &     20\\
$ X300^{  }$  & $ abcdbceeda $  &    29  &  -9.4327~( 194) &     28\\
$ X301^{  }$  & $ abcdbdaeec $  &    31  &  -1.3351~(  81) &     22\\
$ X302^{  }$  & $ abcdbdeeca $  &    59  &  -1.8294~( 223) &     30\\
$ X303^{  }$  & $ abcdbeadec $  &     2  &   0.3341~(   7) &      2\\
$ X304^{  }$  & $ abcdbeaecd $  &     5  &  -0.3397~(  16) &     10\\
$ X305^{  }$  & $ abcdbeaedc $  &     5  &   0.4715~(  14) &      2\\
$ X306^{  }$  & $ abcdbeceda $  &    23  &   0.1228~(  55) &     20\\
$ X307^{  }$  & $ abcdbedeca $  &    47  &  -0.3071~(  59) &     21\\
$ X308^{  }$  & $ abcdbeeadc $  &     6  &   1.8122~(  22) &     20\\
$ X309^{  }$  & $ abcdbeecda $  &    26  &  -4.2448~( 173) &     20\\
$ X310^{  }$  & $ abcdbeedca $  &    50  &   0.2490~( 191) &     21\\
$ X311^{  }$  & $ abcdcabeed $  &    15  &  -0.5291~(  58) &     12\\
$ X312^{  }$  & $ abcdcadeeb $  &    31  &  -1.2454~( 139) &     14\\
$ X313^{  }$  & $ abcdcaebed $  &    11  &   0.9660~(  38) &      4\\
$ X314^{  }$  & $ abcdcaedeb $  &    23  &   0.8266~(  29) &     10\\
$ X315^{  }$  & $ abcdcaeebd $  &    13  &  -1.3728~(  43) &     20\\
$ X316^{  }$  & $ abcdcaeedb $  &    25  &   0.0094~(  39) &     12\\
$ X317^{  }$  & $ abcdcbeeda $  &    59  &   1.4535~( 221) &     23\\
$ X318^{dd}$  & $ abcdcdaeeb $  &    62  &  -8.7568~( 343) &     59\\
$ X319^{  }$  & $ abcdcdeaeb $  &    47  &   0.6801~( 179) &     25\\
$ X320^{  }$  & $ abcdceadeb $  &    11  &   0.5627~(  17) &     10\\
$ X321^{  }$  & $ abcdceaedb $  &    23  &  -0.9005~(  26) &     10\\
$ X322^{  }$  & $ abcdcedaeb $  &    23  &   0.9338~(  23) &      2\\
$ X323^{  }$  & $ abcdceeadb $  &    25  &  -0.0053~(  40) &     12\\
$ X324^{  }$  & $ abcdceedab $  &    53  &  -8.8058~( 243) &     23\\
$ X325^{  }$  & $ abcdceedba $  &   107  &  11.5958~( 343) &     51\\
$ X326^{  }$  & $ abcddabeec $  &    17  &  -9.0047~( 145) &     24\\
$ X327^{  }$  & $ abcddaceeb $  &    33  &   1.5517~( 229) &     29\\
$ X328^{  }$  & $ abcddaebec $  &    13  &  -0.2781~(  42) &     20\\
$ X329^{  }$  & $ abcddaeceb $  &    25  &  -0.9627~(  67) &     11\\
$ X330^{  }$  & $ abcddaeebc $  &    15  &  -4.9591~(  88) &     14\\
$ X331^{  }$  & $ abcddaeecb $  &    27  &   4.7241~( 127) &     25\\
$ X332^{  }$  & $ abcddbaeec $  &    33  &   3.0539~( 161) &     25\\
$ X333^{dd}$  & $ abcddbeeca $  &    65  &   6.8088~( 341) &     49\\
$ X334^{dd}$  & $ abcddcaeeb $  &    47  &   5.1727~( 340) &     23\\
$ X335^{  }$  & $ abcddceaeb $  &    37  &  -2.0294~( 132) &     25\\
$ X336^{  }$  & $ abcddeabec $  &     6  &  -0.7685~(  20) &     20\\
$ X337^{  }$  & $ abcddeaceb $  &    12  &  -1.2039~(  32) &     20\\
$ X338^{  }$  & $ abcddeaebc $  &    13  &  -1.8505~(  38) &     20\\
$ X339^{  }$  & $ abcddeaecb $  &    25  &   0.4111~(  40) &     12\\
$ X340^{  }$  & $ abcddebeca $  &    53  &  -2.1543~( 202) &     25\\
$ X341^{  }$  & $ abcddecaeb $  &    24  &   1.7815~(  33) &     20\\
$ X342^{dd}$  & $ abcddeeacb $  &    27  &   2.6063~( 125) &      0\\
$ X343^{  }$  & $ abcdeabced $  &     2  &   3.8873~(  30) &      6\\
$ X344^{  }$  & $ abcdeabdce $  &     2  &   3.4223~(  18) &      6\\
$ X345^{  }$  & $ abcdeabdec $  &     2  &  -1.0075~(  18) &      4\\
$ X346^{  }$  & $ abcdeabecd $  &     2  &   0.2864~(  20) &      6\\
$ X347^{  }$  & $ abcdeabedc $  &     2  &  -2.6846~(  21) &      6\\
$ X348^{  }$  & $ abcdeacbed $  &     2  &  -0.4899~(  15) &      4\\
$ X349^{  }$  & $ abcdeacdeb $  &     5  &   2.0800~(  36) &      2\\
$ X350^{  }$  & $ abcdeacebd $  &     2  &   1.4643~(  11) &      4\\
$ X351^{  }$  & $ abcdeacedb $  &     5  &   0.2554~(  20) &      2\\
$ X352^{  }$  & $ abcdeadbec $  &     2  &  -0.1260~(   8) &      2\\
$ X353^{  }$  & $ abcdeadceb $  &     5  &   0.1950~(  16) &      2\\
$ X354^{  }$  & $ abcdeadebc $  &     5  &  -2.0503~(  20) &      2\\
$ X355^{  }$  & $ abcdeadecb $  &    11  &  -1.0738~(  25) &      2\\
$ X356^{  }$  & $ abcdeaebcd $  &     5  &   2.0684~(  24) &     10\\
$ X357^{  }$  & $ abcdeaebdc $  &     5  &   0.3746~(  16) &      2\\
$ X358^{  }$  & $ abcdeaecbd $  &     5  &   0.0463~(  16) &      2\\
$ X359^{  }$  & $ abcdeaecdb $  &    11  &  -0.1396~(  17) &     10\\
$ X360^{  }$  & $ abcdeaedbc $  &    11  &  -0.4604~(  37) &      2\\
$ X361^{  }$  & $ abcdeaedcb $  &    23  &   2.5600~(  26) &     10\\
$ X362^{  }$  & $ abcdebadec $  &     2  &  -0.5714~(  12) &      4\\
$ X363^{  }$  & $ abcdebaecd $  &     2  &  -2.3442~(  19) &      4\\
$ X364^{  }$  & $ abcdebaedc $  &     2  &   2.3957~(  18) &      4\\
$ X365^{  }$  & $ abcdebceda $  &    11  &   0.4177~(  30) &     20\\
$ X366^{  }$  & $ abcdebdeca $  &    23  &   5.6759~(  43) &     20\\
$ X367^{  }$  & $ abcdebeadc $  &     5  &  -0.7176~(  12) &     10\\
$ X368^{  }$  & $ abcdebecda $  &    23  &  -0.3404~(  45) &     20\\
$ X369^{  }$  & $ abcdebedca $  &    47  &  -3.3812~(  59) &     21\\
$ X370^{  }$  & $ abcdecadeb $  &     5  &  -1.4763~(  12) &     10\\
$ X371^{  }$  & $ abcdecaedb $  &     5  &   0.0045~(  10) &      2\\
$ X372^{  }$  & $ abcdecdaeb $  &    11  &  -1.2900~(  33) &      2\\
$ X373^{  }$  & $ abcdeceadb $  &    23  &   0.5851~(  24) &      2\\
$ X374^{  }$  & $ abcdecedab $  &    47  &   0.9188~( 266) &     18\\
$ X375^{  }$  & $ abcdecedba $  &    89  &   1.0991~( 163) &     25\\
$ X376^{  }$  & $ abcdedabec $  &     5  &   1.0484~(  16) &      2\\
$ X377^{  }$  & $ abcdedaceb $  &    11  &   0.4264~(  27) &      2\\
$ X378^{  }$  & $ abcdedaebc $  &    11  &   1.3196~(  21) &      2\\
$ X379^{  }$  & $ abcdedaecb $  &    23  &  -0.3201~(  17) &     10\\
$ X380^{  }$  & $ abcdedbeca $  &    47  &  -1.0268~(  48) &     21\\
$ X381^{  }$  & $ abcdedcaeb $  &    23  &   1.0861~(  29) &      2\\
$ X382^{  }$  & $ abcdedeacb $  &    41  &  -1.7712~(  80) &     21\\
$ X383^{  }$  & $ abcdeeabdc $  &     6  &  -4.8034~(  22) &     20\\
$ X384^{  }$  & $ abcdeeacdb $  &    12  &   1.9266~(  31) &     20\\
$ X385^{  }$  & $ abcdeeadbc $  &    12  &  -0.7427~(  19) &     20\\
$ X386^{  }$  & $ abcdeeadcb $  &    24  &   0.6887~(  38) &     11\\
$ X387^{  }$  & $ abcdeebdca $  &    50  &   1.9508~( 152) &     21\\
$ X388^{  }$  & $ abcdeecadb $  &    24  &  -0.4349~(  40) &     20\\
$ X389^{  }$  & $ abcdeedacb $  &    30  &  -0.0433~(  68) &     25\\
%================================================================
 \end{longtable*}
\endgroup

\renewcommand{\baselinestretch}{0.95}
\begin{table*}
\caption{
Residual renormalization constants used to calculate
$a_e^{(10)} [\text{Set~V}]$. 
Notations are those of Eq.~(\ref{a_Vr}).
 \label{Table:residual_const}
}
\hfill
\begin{minipage}[t][][b]{.49\textwidth}
\begin{ruledtabular}
\begin{tabular}{@{\hskip1em}l@{}d@{\hskip1em}}
\multicolumn{1}{c}{Integral} &
\multicolumn{1}{c}{Value (Error)} \\
\hline
    $ \Delta M_{10}$  &        3.468~(336)          \\
    $ \Delta M_{8}$   &        1.738~12~(85)        \\
    $ \Delta M_{6}$   &        0.425~8135~(30)      \\
    $ \Delta M_{4}$   &        0.030~833~612\cdots  \\
    $  M_{2}$         &        0.5                  \\
    $ \Delta L\!B_{8}$  &        2.0504~(86)        \\ 
    $ \Delta L\!B_{6}$  &        0.100~801~(43)     \\  
    $ \Delta L\!B_{4}$  &        0.027~9171~(61)    \\ 
    $ \Delta L\!B_{2}$  &        0.75               \\
\end{tabular}
\end{ruledtabular}
\end{minipage}
%
%\hskip-.5em
\hfill
\begin{minipage}[t][][b]{.49\textwidth}
\begin{ruledtabular}
\begin{tabular}{@{\hskip1em}l@{}d@{\hskip1em}}
\multicolumn{1}{c}{Integral} &
\multicolumn{1}{c}{Value (Error)} \\
\hline
    $ \Delta L_{4^*}$ &       -0.459~051~(62)         \\ 
    $ \Delta L_{2^*}$ &       -0.75                   \\
    $ \Delta dm_{6}$  &       -2.340~815~(55)         \\ 
    $ \Delta dm_{4}$  &        1.906~3609~(90)        \\
    $ \Delta dm_{2^*}$&       -0.75                   \\
\\
\\
\\
\\
\end{tabular}
\end{ruledtabular}
\end{minipage}
\hfill
\end{table*}

\renewcommand{\baselinestretch}{1.0}

%----------------------------------------------------------------

%================================================================
\begin{acknowledgments}
We thank Dr.~M.~Steinhauser for his communication concerning the eighth-order 
mass-dependent term. We also appreciate communications with Dr.~P.~Mohr. 
This work is supported in part by 
the Japan Society for the Promotion of Science (JSPS) 
Grant-in-Aid for Scientific Research (C)20540261 and (C)23540331.
T.~K.'s work is supported in part by the U.~S.~National Science Foundation
under Grant No.~NSF-PHY-0757868.
T.~K.~thanks RIKEN for the hospitality extended to him
while a part of this work was carried out.
Numerical calculations are conducted on 
the RIKEN Super Combined Cluster (RSCC) and 
the RIKEN Integrated Cluster of Clusters (RICC) 
supercomputing systems.
\end{acknowledgments}
%----------------------------------------------------------------

%\newpage

%================================================================
\appendix
%----------------------------------------------------------------

\section{{\it K}-operation, {\it R}-subtraction, and (modified) {\it I}-operation on the diagram {\rm X253}}
\label{sec:appendixX253}

\begin{figure}[tb]
\includegraphics[scale=.5]{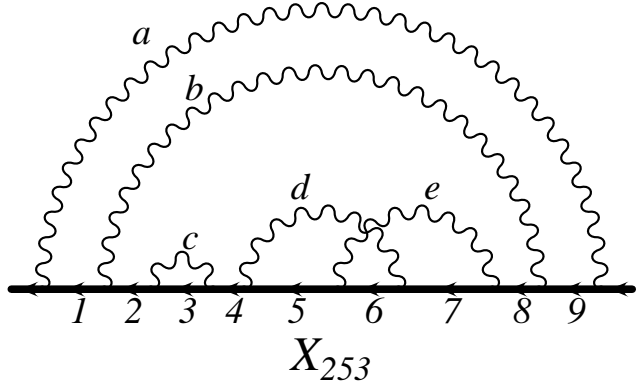}
\caption{
Diagram {\rm X253}.
\label{fig:X253}
}
\end{figure}

This Appendix is devoted to the discussion of the
diagram {\rm X253} shown in Fig. \ref{fig:X253}. 
We describe in some detail
the relation between the standard on-shell renormalization
and the renormalization method
adopted by {\sc gencode}{\it N} based on the {\it K}-operation, 
{\it R}-subtraction, and {\it I}-subtraction, 
choosing {\rm X253} as an example.
Actually, {\rm X253} and another diagram X256 are not entirely typical
in the sense that they require a slight modification of one of
the {\it I}-subtraction operations encoded in {\sc gencode}{\it N}.
The reason why this modification is required and its resolution will be discussed in detail.

In this Appendix, we adopt the following notations. 
Internal lepton lines are numbered 
{\sl 1, 2, 3, 4, 5, 6, 7, 8, 9} 
from left to right,
and internal photon lines are numbered 
{\sl a, b, c, d, e} 
as shown in Fig. 
\ref{fig:X253}.
Subdiagrams are represented by the set of indices enclosed in braces.
The subtraction operators are labelled by the indices of lepton lines 
of the subdiagrams: 
For example, the {\it K}-operation applied to the self-energy subdiagram 
{\sl \{3; c\} }
is denoted as ${\mathbb K}_{3}$. 
The {\it R}-subtraction applied to the self-energy-like subdiagram 
{\sl \{5 6 7; d e\} } is denoted as ${\mathbb R}_{567}$. 
The {\it I}-subtraction applied to the self-energy-like subdiagram
{\sl \{2 3 4 5 6 7 8; b c d e\} } is denoted as ${\mathbb I}_{19}$
using the indices of the residual diagram {\sl \{1 9; a\} }
which is obtained by reducing the subdiagram to a point. 
For the nested {\it I}-subtractions applied to subdiagrams 
$\mathcal{S}_1$ and $\mathcal{S}_2$ 
where $\mathcal{S}_1 \supset \mathcal{S}_2$, 
the operators are labelled by the indices in the reduced subdiagrams 
$\mathcal{G}/\mathcal{S}_1$ and $\mathcal{S}_1/\mathcal{S}_2$, respectively. 
Other cases are denoted in a similar manner accordingly. 

\subsection{Standard renormalization}

The diagram {\rm X253} has UV divergences arising from the following subdiagrams:
{\sl \{3; c\}}, 
{\sl \{5 6 7; d e\}}, 
{\sl \{5 6; d\}}, 
{\sl \{6 7; e\}}, 
{\sl \{2 3 4 5 6 7 8; b c d e\}}.
Recalling that Fig.~\ref{fig:X253} actually represents 
the sum of nine vertex diagrams containing various subdiagrams of 
vertex type and self-energy type, 
we can write the standard renormalization of {\rm X253} as follows:
%
%----------------------------------------------------------------
\begin{eqnarray}
a_{\rm X253} &=&M_{\rm X253} 
          + M_{30}~ (  - 2~L_{2} ) 
          + M_{42}~ (  - B_{2} )
          + M_{42(2^*)}~ (  - dm_{2} )
\nonumber \\
         &+&M_{6b}~ ( -B_{4a}+ 4~L_{2}~B_{2}  )
          + M_{6b(2^*)}~ ( -dm_{4a} + 4~L_{2}~dm_{2} )
\nonumber \\
         &+&M_{4b}~ \{ 
                 B_{4a}~B_{2} - 2~L_2~ (B_{2})^2
                     \} 
          + M_{4b(2^*)}~ \{ 
                dm_{4a} B_2 + dm_{2}~(B_{4a}- 4~L_{2}~B_{2}) 
                         \} 
\nonumber \\
          &+&M_{4b(2^{**})}~ dm_{2}~ (  dm_{4a} - 2~L_{2}~dm_{2} )
\nonumber \\
      &+&M_{2}~ [ 
               - B_{16} 
               + 2~B_{6a}~L_{2}
               + B_{6c}~B_{2} 
               + B_{4a}~\{ B_{4b} - (B_{2})^2 \}
\nonumber \\
              && - 4~B_{4b}~L_{2}~B_{2} + 2~L_2~ (B_{2})^3 
                 + dm_{4a}~(  B_{4b(1^*)} - B_{2}~B_{2^*} ) 
                ]
\nonumber \\
      &+&M_{2}~dm_{2}~  \{ 
                       B_{6c(1^*)} 
                     - 4~L_{2}~B_{4b(1^*)} - B_{2^{**}}~dm_{4a} 
                     - B_{2^*}~B_{4a} 
\nonumber \\
                    && + 2~L_{2}~( dm_{2}~B_{2^{**}} +2~B_{2}~B_{2^*} ) 
                      \} 
\nonumber \\
      &+& M_{2^*}~ dm_{2}~ [ 
               dm_{6c(1^*)}  - B_{4a} (  dm_{2^*} +  B_{2}  ) 
             - dm_{2^{**}}~dm_{4a} - 4~L_{2}~dm_{4b(1^*)}  
\nonumber \\
             &&+ 2~L_{2} \{  (B_{2})^2 + 2~B_{2}~dm_{2^*} + dm_{2}~dm_{2^{**}}
                        \}
                              ]
\nonumber \\
      &+& M_{2^*}~ \{
                 - dm_{16} 
                 + 2~dm_{6a}~L_2 
                 + dm_{6c}~B_2 
\nonumber \\ 
              && + dm_{4a}~(dm_{4b(1^*)}-B_{2}~dm_{2^*}) 
                 + dm_{4b}~(B_{4a}-4~L_{2}~B_{2} ) 
                   \} 
\label{appA1}
\end{eqnarray}
%----------------------------------------------------------------
%
\noindent
where the suffixes 16, 30, and 42  
are those identifying eighth-order subdiagrams \cite{Kinoshita:2005sm}.
Suffixes $6a$, $6b$, $6c$ refer to the sixth-order subdiagrams, and 
$4a$ ,$4b$ refer to the fourth-order subdiagrams.
Symbols $(i^*)$ in the suffixes refer to the diagrams derived by 
insertion of a two-point vertex in the lepton line $i$. $(i^{**})$ 
corresponds to two insertions of vertices. For second-order diagrams, 
the parentheses and the index $i$ are omitted for simplicity. 
See Ref.~\cite{Kinoshita:2005sm} for
the explanation of other notations.
All terms on the right-hand side of (\ref{appA1}) 
contain UV divergent parts.
Thus some regularization is assumed.

\subsection{Separation of UV divergences by {\it K}-operation}

The first step is to separate UV-divergent parts of
all terms on the right-hand side of (\ref{appA1}) 
from their UV-finite parts.
We carry out this separation by means of {\it K}-operation
starting with $M_{\rm X253}$.

$M_{\rm X253}$ has no overall UV divergence.
It has only UV divergences from some subdiagrams.
The UV-divergence-free part $M_{\rm X253}^{\rm R}$ of $M_{\rm X253}$
is defined by the {\it K}-operations as 
\begin{equation}
M_{\rm X253}^R = \sum_{f\in\mathfrak{F}({\cal G})}
        \left[
                \prod_{{\cal S}_i\in f} (-\mathbb{K}_{{\cal S}_i})
        \right]
        M_{\rm X253},
\end{equation}
where the sum is over 24 forests constructed from the five subdiagrams 
including the empty forest.
Note that the forest corresponding to ${\mathbb K}_{56} {\mathbb K}_{67}$
is absent since the subdiagrams {\sl \{5 6; d\}} and {\sl \{6 7; e\}} 
overlap each other. 

Carrying out the {\it K}-operations explicitly (see Sec.~\ref{subsec:UVsubtractionterms}), 
and rewriting the result as an expression of $M_{\rm X253}$,  we obtain
%
%----------------------------------------------------------------
\begin{eqnarray}
M_{\rm X253} &=& M_{\rm X253}^{\rm R} 
       + M_{30} ~ ( 2~L_{2}^{\rm UV} )
       + M_{42} ~ ( B_{2}^{\rm UV} )
       + M_{42(2^*)} ~ ( dm_{2}^{\rm UV} )
\nonumber \\
      &+&M_{6b} ~ (   B_{4a}^{\rm UV} - 4~L_{2}^{\rm UV}~B_{2}^{\rm UV}  )
       + M_{6b(2^*)} ~ ( dm_{4a}^{\rm UV} - 4~L_{2}^{\rm UV}~dm_{2}^{\rm UV} )
\nonumber \\
      &+&M_{4b} ~ \{  - B_{4a}^{\rm UV}~B_{2}^{\rm UV}  + 2~L_{2}^{\rm UV}~(B_{2}^{\rm UV})^2 \} 
\nonumber \\
      &+& M_{4b(2^*)} ~ \{ - dm_{4a}^{\rm UV}~B_{2}^{\rm UV}  
                    + dm_{2}^{\rm UV}~(-B_{4a}^{\rm UV} + 4~B_{2}^{\rm UV}~L_{2}^{\rm UV} )  \}
\nonumber \\
      &+&M_{4b(2^{**})} ~ dm_{2}^{\rm UV}~(  - dm_{4a}^{\rm UV} + 2~L_{2}^{\rm UV}~dm_{2}^{\rm UV} )
\nonumber \\
      &+&M_{2} ~ [ 
                  B_{16}^{\rm UV} 
                - 2~B_{6a}^{\rm UV}~L_{2}^{\rm UV}
                - B_{6c(1^\prime)}^{\rm UV}~B_{2}^{\rm UV} 
                + B_{4a}^{\rm UV}~(- B_{4b(1^\prime)}^{\rm UV} 
                           + B_{2}^{\rm UV}~B_{2^{\prime\prime}}^{\rm UV} 
                            )
\nonumber \\
            &&
                + 4 ~B_{4b(1^\prime)}^{\rm UV} ~L_{2}^{\rm UV}~B_{2}^{\rm UV}
                - 2~L_{2}^{\rm UV}~(B_{2}^{\rm UV})^2~B_{2^{\prime\prime}}^{\rm UV} 
                  ]
\nonumber \\
      &+&M_{2^*} [ dm_{2}^{\rm UV} \{  
                - ~dm_{6c(1^*)}^{\rm UV}
                + B_{4a}^{\rm UV}~dm_{2^{*\prime}}^{\rm UV} 
                + 4~L_{2}^{\rm UV}~dm_{4b(1^*)}^{\rm UV} 
                - 4~L_{2}^{\rm UV}~B_{2}^{\rm UV}~dm_{2^{*\prime}}^{\rm UV}
                           \}
\nonumber \\
             &&   + dm_{2^{\prime\prime}}^{\rm UV}  \{
                     B_{4a}^{\rm UV}~B_{2}^{\rm UV} 
                   - 2~L_{2}^{\rm UV}~(B_{2}^{\rm UV})^2
                            \}
                  ]
\nonumber \\
       &+& M_{2^*} ~ \{ 
                  dm_{16}^{\rm UV} 
                - 2~dm_{6a}^{\rm UV}~L_{2}^{\rm UV} 
                - dm_{6c(1^\prime)}^{\rm UV} ~B_{2}^{\rm UV}
\nonumber \\
              &&  + dm_{4a}^{\rm UV}~( -dm_{4b(1^*)}^{\rm UV} + B_{2}^{\rm UV}~dm_{2^{*\prime}}^{\rm UV} ) 
                + dm_{4b(1^\prime)}^{\rm UV} ( -B_{4a}^{\rm UV} + 4~L_{2}^{\rm UV}~B_{2}^{\rm UV}  )
                   \}.
\label{MX253UV}
\end{eqnarray}
%----------------------------------------------------------------
%
Here, the symbols with a primed suffix $i^\prime$ represents a quantity 
obtained by differentiating the amplitude with respect to $z_i$.
For the second order case the index is omitted for simplicity. 
See Ref.~\cite{Kinoshita:1990} for further explanations. 

The next step is to substitute (\ref{MX253UV}) in (\ref{appA1}).
Since the result of substitution 
contains eighth-order terms $M_{30}$, $M_{42}$, etc., which are UV-divergent,
we must substitute them by the {\it K}-operation results
of $M_{30}$, $M_{42}$, etc., listed below:
%
%----------------------------------------------------------------
\begin{eqnarray} 
M_{30}&=&M_{30}^{\rm R}
       +2dm_{2}^{\rm UV}  M_{6b(2^*)} +2B_{2}^{\rm UV}  M_{6b}
       +dm_{6a}^{\rm UV}  M_{2^*}+B_{6a}^{\rm UV}  M_{2}
\nonumber \\
      &-&dm_{2}^{\rm UV}  (dm_{2}^{\rm UV}  M_{4b(2^{**})} +B_{2}^{\rm UV}  M_{4b(2^*)})
      -B_{2}^{\rm UV}  (dm_{2}^{\rm UV}  M_{4b(2^*)}+B_{2}^{\rm UV}  M_{4b})
\nonumber \\
      &-&2dm_{2}^{\rm UV}  dm_{4b(1^*)}^{\rm UV}  M_{2^*}
       - 2B_{2}^{\rm UV}  (dm_{4b(1^\prime)}^{\rm UV}  M_{2^*}
                     +B_{4b(1^\prime)}^{\rm UV}  M_{2})
\nonumber \\
      &+&2dm_{2}^{\rm UV}  B_{2}^{\rm UV}  dm_{2^{* \prime}}^{\rm UV}  M_{2^*}
      +(B_{2}^{\rm UV})^2(dm_{2^{\prime\prime}}^{\rm UV}  M_{2^*}
                       +B_{2^{\prime\prime}}^{\rm UV}  M_{2}),
\end{eqnarray}
\begin{eqnarray} 
M_{42}&=&M_{42}^{\rm R}+2L_{2}^{\rm UV}  M_{6b}
         +dm_{4a}^{\rm UV}  M_{4b(2^*)}+B_{4a}^{\rm UV}  M_{4b}
         +dm_{6c}^{\rm UV}  M_{2^*}+B_{6c}^{\rm UV}  M_{2}
\nonumber \\
        &-&2L_{2}^{\rm UV}  (dm_{2}^{\rm UV}  M_{4b(2^*)}+B_{2}^{\rm UV}  M_{4b})
         -2L_{2}^{\rm UV}  (dm_{4b}^{\rm UV}  M_{2^*}+B_{4b}^{\rm UV}  M_{2})
\nonumber \\
        &-&dm_{4a}^{\rm UV}  dm_{2^*}^{\rm UV}  M_{2^*}
         -B_{4a}^{\rm UV}  (dm_{2^\prime}^{\rm UV}  M_{2^*} 
                 +B_{2^\prime}^{\rm UV}  M_{2})
\nonumber \\
        & +&2L_{2}^{\rm UV}  dm_{2}^{\rm UV}  dm_{2^*}^{\rm UV}  M_{2^*}
         +2L_{2}^{\rm UV}  B_{2}^{\rm UV}  (dm_{2^\prime}^{\rm UV}  M_{2^*}
                +B_{2^\prime}^{\rm UV}  M_{2}),
\end{eqnarray}
\begin{eqnarray} 
B_{16}&=& B_{16}^{\rm UV} + B_{16}^{\rm R}
        +2L_{2}^{\rm UV}  \widetilde{B_{6a}}
        +dm_{2}^{\rm UV}  B_{6c(1^*)}+B_{2}^{\rm UV}  \widetilde{B_{6c(1^\prime)}}
\nonumber \\
       &+&(dm_{4a}^{\rm UV}-4 dm_{2}^{\rm UV}  L_{2}^{\rm UV})  B_{4b(1^*)}
        +(B_{4a}^{\rm UV}-4 L_{2}^{\rm UV}  B_{2}^{\rm UV})  \widetilde{B_{4b(1^\prime)}}
\nonumber \\
       & +&(-B_{2}^{\rm UV}  dm_{4a}^{\rm UV} -dm_{2}^{\rm UV}  B_{4a}^{\rm UV} +4 dm_{2}^{\rm UV}  L_{2}^{\rm UV}  B_{2}^{\rm UV})  B_{2^{* \prime}}
\nonumber \\
       &-&dm_{2}^{\rm UV}  (dm_{4a}^{\rm UV}-2L_{2}^{\rm UV}  dm_{2}^{\rm UV})  B_{2^{**}}
        -B_{2}^{\rm UV}  (B_{4a}^{\rm UV}-2L_{2}^{\rm UV}  B_{2}^{\rm UV})  \widetilde{B_{2^{\prime\prime}}},
\end{eqnarray}
\begin{eqnarray} 
 dm_{16}&=&dm_{16}^{\rm UV} + dm_{16}^{\rm R}
      + 2L_{2}^{\rm UV}  \widetilde{dm_{6a}}
      + dm_{2}^{\rm UV}  \widetilde{dm_{6c(1^*)}} 
          +B_{2}^{\rm UV}  \widetilde{dm_{6c(1^\prime)}}
\nonumber \\
     &+&(dm_{4a}^{\rm UV} -4 dm_{2}^{\rm UV}  L_{2}^{\rm UV})  \widetilde{dm_{4b(1^*)}} 
      + (B_{4a}^{\rm UV}-4 L_{2}^{\rm UV}  B_{2}^{\rm UV})  \widetilde{dm_{4b(1^\prime)}}
\nonumber \\
     & +& (-B_{2}^{\rm UV}  dm_{4a}^{\rm UV} -dm_{2}^{\rm UV}  B_{4a}^{\rm UV} 
                 +4 dm_{2}^{\rm UV}  L_{2}^{\rm UV}  B_{2}^{\rm UV})  \widetilde{dm_{2^{* \prime}}}
\nonumber \\
     &-&dm_{2}^{\rm UV}  (dm_{4a}^{\rm UV}-2L_{2}^{\rm UV}  dm_{2}^{\rm UV})  dm_{2^{**}} 
      - B_{2}^{\rm UV}  (B_{4a}^{\rm UV} -2 L_{2}^{\rm UV}  B_{2}^{\rm UV})  \widetilde{dm_{2^{\prime\prime}}},
\label{otherUVterms}
\end{eqnarray}
and so on, where $\widetilde{{B}_{6a}} \equiv B_{6a} - B_{6a}^{\rm UV}$, etc.
Note that $M_{30}$ and $M_{42}$ have 
UV divergences coming from subdiagrams but no overall UV divergences
whereas renormalization constants
 $B_{16}$ and $dm_{16}$ have overall UV divergences.
     
Since the result of substitution of (\ref{otherUVterms})
still contains $M_{6a}$, etc., which have UV-divergent subdiagrams,
it is necessary to separate their UV-divergent parts using
\begin{eqnarray}
M_{6b}&=&M_{6b}^{\rm R} +dm_{2}^{\rm UV}  M_{4b(2^*)}+B_{2}^{\rm UV}  M_{4b}
         +(dm_{4b}^{\rm UV}-dm_{2}^{\rm UV}  dm_{2^*}^{\rm UV})  M_{2^*}
\nonumber \\
       &+&B_{4b}^{\rm UV}  M_{2}
        -B_{2}^{\rm UV}  (dm_{2^\prime}^{\rm UV}  M_{2^*}
               +B_{2^\prime}^{\rm UV}  M_{2}),
\nonumber \\
M_{6b(2^*)} &=& M_{6b(2^*)}^{\rm R}
         + dm_{2}^{\rm UV}  M_{4b(2^{**})} + B_{2}^{\rm UV}  M_{4b(2^*)}
         + dm_{4b(1^*)}^{\rm UV}  M_{2^*}  
         - dm_{2^{*\prime}}^{\rm UV}  B_{2}^{\rm UV}  M_{2^*},
\label{Eq:M6b2s}
\end{eqnarray}
\begin{eqnarray}
B_{6a}&=& B_{6a}^{\rm UV} + B_{6a}^{\rm R}
     +2 (dm_{2}^{\rm UV} B_{4b(1^*)}
      +B_{2}^{\rm UV} \widetilde{B_{4b(1^\prime)}})
     -dm_{2}^{\rm UV} (dm_{2}^{\rm UV} B_{2^{**}}+B_{2}^{\rm UV} B_{2^*})
\nonumber \\
   &-&B_{2}^{\rm UV} (dm_{2}^{\rm UV} B_{2^*}+B_{2}^{\rm UV} \widetilde{B_{2^{\prime \prime}}}),
\nonumber \\
B_{6c}&=&B_{6c}^{\rm UV}+B_{6c}^{\rm R}+2 L_{2}^{\rm UV} \widetilde{B_{4b}}+dm_{4a}^{\rm UV} B_{2^*}+B_{4a}^{\rm UV} \widetilde{B_{2^\prime}}
\nonumber \\
       &-&2 L_{2}^{\rm UV} (dm_{2}^{\rm UV} B_{2^*}+B_{2}^{\rm UV} \widetilde{B_{2^\prime}}),
\end{eqnarray}
\begin{eqnarray}
dm_{6a}&=&dm_{6a}^{\rm UV}+dm_{6a}^{\rm R}+2 (dm_{2}^{\rm UV} \widetilde{dm_{4b(1^*)}}
       +B_{2}^{\rm UV} \widetilde{dm_{4b(1^\prime)}})
       -dm_{2}^{\rm UV} (dm_{2}^{\rm UV} dm_{2^{**}}+B_{2}^{\rm UV} \widetilde{dm_{2^{* \prime}}})
\nonumber \\
      & -&B_{2}^{\rm UV} (dm_{2}^{\rm UV}\widetilde{dm_{2^{* \prime}}}
       +B_{2}^{\rm UV} \widetilde{dm_{2^{\prime\prime}}}),
\nonumber \\
dm_{6c}&=&dm_{6c}^{\rm UV} + dm_{6c}^{\rm R}+2 L_{2}^{\rm UV} \widetilde{dm_{4b}}
       +dm_{4a}^{\rm UV} \widetilde{dm_{2^*}}+B_{4a}^{\rm UV} \widetilde{dm_{2^\prime}}
\nonumber \\
     &-&2 L_{2}^{\rm UV} (dm_{2}^{\rm UV} \widetilde{dm_{2^*}}+B_{2}^{\rm UV} \widetilde{dm_{2^\prime}}),
\end{eqnarray}
followed by
\begin{eqnarray}
 M_{4b} &=& M_{4b}^{\rm R} + dm_2^{\rm UV} M_{2^*} + B_{2}^{\rm UV} M_{2},
\nonumber \\
 M_{4b(2^*)} &=& M_{4b(2^*)}^{\rm R} + dm_{2^*}^{\rm UV} M_{2^*},
\\
 B_{4a}&=&B_{4a}^{\rm UV}+B_{4a}^{\rm R}+2 L_2^{\rm UV} B_{2}^{\rm R},
\nonumber \\
 B_{4b}&=&B_{4b}^{\rm UV}+B_{4b}^{\rm R}+dm_{2}^{\rm UV} B_{2^*}+B_{2}^{\rm UV} B_{2^\prime}^{\rm R},
\nonumber \\
B_{4b(1^*)} &=& B_{4b(1^*)}^{\rm R} + dm_2^{\rm UV}  B_{2^{**}} + B_{2}^{\rm UV} B_{2^*},
\nonumber \\
dm_{4a}&=&dm_{4a}^{\rm UV}+dm_{4a}^{\rm R},
\nonumber \label{Eq:dm4a} \\
dm_{4b}&=&dm_{4b}^{\rm UV}+dm_{4b}^{\rm R}
           +dm_2^{\rm UV} dm_{2^*}^{\rm R}+B_2^{\rm UV} dm_{2^\prime}^{\rm R},
\nonumber \\
dm_{4b(1^*)}  &=& dm_{4b(1^*)}^{\rm UV}+ dm_{4b(1^*)}^{\rm R} + dm_2^{\rm UV} dm_{2^{**}}                + B_2^{\rm UV} \widetilde{dm_{2^{*\prime}}},
\label{Eq:dm4b1s} \\
L_2 &=& L_2^{\rm UV} + L_2^{\rm R} ,
\nonumber \\
B_2 &=& B_2^{\rm UV} + B_2^{\rm R} ,
\nonumber \\
dm_{2^*} &=& dm_{2^*}^{\rm UV} + dm_{2^*}^{\rm R}.
\end{eqnarray}
%----------------------------------------------------------------

After all UV divergences are separated out by successive
{\it K}-operations, we can at last express $a_{\rm X253}$ in terms 
of UV-finite quantities only:
%
%----------------------------------------------------------------
\begin{eqnarray}
a_{\rm X253} 
       & = & M_{\rm X253}^{\rm R}
        +  M_{30}^{\rm R} ~ (  - 2~L_{2}^{\rm R} )
        +  M_{42}^{\rm R} ~ (  - B_{2}^{\rm R} )
\nonumber \\
       & + & M_{6b}^{\rm R} ~ (  - B_{4a}^{\rm R} + 4~L_{2}^{\rm R}~B_{2}^{\rm R} )
        +  M_{6b(2^*)}^{\rm R} ~ ( - dm_{4a}^{\rm R} ) 
\nonumber \\
       & + & M_{4b}^{\rm R} ~ \{  B_{4a}^{\rm R}~B_{2}^{\rm R} - 2~L_{2}^{\rm R}~(B_{2}^{\rm R})^2 \} 
        +  M_{4b(2^*)}^{\rm R} ~ dm_{4a}^{\rm R}~B_{2}^{\rm R} 
\nonumber \\
       & + & M_{2} ~ [  - B_{16}^{\rm R} + 2~B_{6a}^{\rm R}~L_{2}^{\rm R} 
                     + B_{6c}^{\rm R}~B_{2}^{\rm R} 
                     + B_{4a}^{\rm R}~\{ B_{4b}^{\rm R} - (B_{2}^{\rm R})^2 \} 
\nonumber \\
                    && - 4~B_{4b}^{\rm R}~L_{2}^{\rm R}~B_{2}^{\rm R} 
                     + 2~L_{2}^{\rm R}~(B_{2}^{\rm R})^3  
                     + dm_{4a}^{\rm R}~(B_{4b(1^*)}^{\rm R} -B_{2^*}~B_{2}^{\rm R})   ] 
\nonumber \\
       & + & M_{2^*}~ \{  - dm_{16}^{\rm R} + 2~dm_{6a}^{\rm R}~L_{2}^{\rm R} 
                          + dm_{6c}^{\rm R}~B_{2}^{\rm R} 
\nonumber \\             
                      &&+ dm_{4a}^{\rm R}
                        ~(dm_{4b(1^*)}^{\rm R}-B_{2}^{\rm R}~dm_{2^*}^{\rm R})  
                        + dm_{4b}^{\rm R}
                         ~(B_{4a}^{\rm R}-4~L_{2}^{\rm R}~B_{2}^{\rm R})   \}.
\label{aX253UVfree}
\end{eqnarray}
%----------------------------------------------------------------
%
Note that (\ref{aX253UVfree}) has exactly the same structure as (\ref{appA1})
but looks simpler because 
$dm_2^{\rm R} \equiv dm_2 - dm_2^{\rm UV} = 0$.

\subsection{Separation of IR divergences by {\it R}-subtraction and {\it I}-subtraction}

The integrands of $M_{\rm X253}^{\rm R}$, etc., are singular at vanishing momenta 
of virtual photons because of vanishing photon mass.
When the integrands are integrated over all momenta, these singularities
give rise to logarithmic IR divergences (if enhanced by
vanishing denominators of two lepton propagators
which are adjacent to the external lines)
or linear IR divergences (if enhanced by three lepton propagators).

To prepare for the numerical integration 
it is necessary to separate the IR-divergent parts
from the IR-finite parts, and integrate only the latter parts.
Since the sum of all diagrams of Set V is gauge-invariant and finite,
IR-divergent parts cancel out when summed over all diagrams of Set V.

As we have discussed in Ref.~\cite{Aoyama:2007bs}
and Sec.~\ref{subsec:IRsubtractionterms}, the IR divergences in the
amplitude $M_{\rm X253}^{\rm R}$ 
can be handled completely by looking at the self-energy-like subdiagram
$\mathcal{S}$ of ${\rm X253}$. They are 
$\mathcal{S}_1$ = {\sl \{2 3 4 5 6 7 8; b c d e\}}, 
$\mathcal{S}_2$ = {\sl \{5 6 7; d e\}}, and 
$\mathcal{S}_3$ = {\sl \{3; c\}}. 
There are two subtraction schemes, {\it R}-subtraction
to deal with the linear IR divergence, and {\it I}-subtraction
 to deal with the logarithmic IR divergence:
\begin{itemize}
\item {\it R}-subtraction annotates $M$ and $dm$ to 
the whole diagram $\mathcal{G}$ 
and some of the subdiagrams $\mathcal{S}$, respectively.  
Following the procedure built into {\sc gencode}{\it N}
{\it R}-subtraction $\mathbb{R}_\mathcal{S}$ 
is applied to the subdiagram $\mathcal{S}$. 
The reduced diagram $\mathcal{G/S}$ gives rise to a magnetic moment amplitude
of lower order.

\item {\it I}-subtraction annotates $I$ and $M$ to the whole diagram $\mathcal{G}$ 
and some of the subdiagrams $\mathcal{S}$, respectively. 
Then {\it I}-subtraction $\mathbb{I}_\mathcal{S}$
is applied to the reduced diagram $\mathcal{G/S}$
and the subdiagram $\mathcal{S}$ gives rise to a magnetic moment amplitude
of lower order.

\item In addition there are cases where {\it R}-subtraction 
and {\it I}-subtraction occur together.

\end{itemize}

The diagram ${\rm X253}$ has 11 annotated forests.
{\sc gencode}{\it N} generates IR subtraction terms as follows, 
where $\mathbb{R}_{2-8}$ is an abbreviation of $\mathbb{R}_{2345678}$: 
\[
\begin{array}{l@{\qquad}l@{\qquad}l}
\text{annotation}
& \text{subtraction}      
& \text{expression} 
\\
\mathcal{G}\rightarrow M, \quad \mathcal{S}_1\rightarrow dm
& {\mathbb R}_{2-8}
& dm_{16}^{\rm R} M_{2^*}
\\
\mathcal{G}\rightarrow M, \quad \mathcal{S}_2\rightarrow dm
& {\mathbb R}_{567}
& dm_{4a}^{\rm R} M_{6b(2^*)}^{\rm R}
\\
\mathcal{G}\rightarrow M, \quad \mathcal{S}_1\rightarrow dm, \quad \mathcal{S}_2\rightarrow dm
& {\mathbb R}_{2-8}{\mathbb R}_{567}
& dm_{4b(1^*)}^{\rm R} dm_{4a}^{\rm R} M_{2^*}
\\
\mathcal{G}\rightarrow I, \quad \mathcal{S}_1\rightarrow M
& {\mathbb I}_{19}
& L_2^{\rm R} M_{16}^{\rm R}
\\
\mathcal{G}\rightarrow I, \quad \mathcal{S}_2\rightarrow M
& {\mathbb I}_{123489}
& L_{6b(2)}^{\rm R} M_{4a}^{\rm R}
\\
\mathcal{G}\rightarrow I, \quad \mathcal{S}_3\rightarrow M
& {\mathbb I}_{12456789}
& L_{42(2)}^{\rm R} M_2
\\
\mathcal{G}\rightarrow I, \quad \mathcal{S}_1\rightarrow I, \quad \mathcal{S}_2\rightarrow M
& {\mathbb I}_{19}{\mathbb I}_{2348}
& L_2^{\rm R} L_{4b1}^{\rm R} M_{4a}^{\rm R}
\\
\mathcal{G}\rightarrow I, \quad \mathcal{S}_1\rightarrow I, \quad \mathcal{S}_3\rightarrow M
& {\mathbb I}_{19}{\mathbb I}_{245678}
& L_2^{\rm R} L_{6c(1)}^{\rm R} M_2
\\
\mathcal{G}\rightarrow I, \quad \mathcal{S}_1\rightarrow M, \quad \mathcal{S}_2\rightarrow dm
& {\mathbb I}_{19}{\mathbb R}_{567}
& L_2^{\rm R} dm_{4a}^{\rm R} M_{4b(1^*)}^{\rm R}
\\
\mathcal{G}\rightarrow I, \quad \mathcal{S}_3\rightarrow M, \quad \mathcal{S}_2 \rightarrow dm
& {\mathbb I}_{12489}{\mathbb R}_{567}
& L_{4b2(2^*)}^{\rm R} dm_{4a}^{\rm R} M_2
\\
\mathcal{G}\rightarrow I, \quad \mathcal{S}_1\rightarrow I, \quad \mathcal{S}_3 \rightarrow M, \quad \mathcal{S}_2\rightarrow dm
& {\mathbb I}_{19}{\mathbb I}_{248}{\mathbb R}_{567}
& L_2^{\rm R} L_{2^*}^{\rm R} dm_{4a}^{\rm R} M_2
\end{array}
\]

In the diagram $M_{\rm X253}^{\rm R}$ one of the linear IR divergences
occurs when the momentum of the outermost photon {\sl a} vanishes.
The self-energy-like subdiagram {\sl \{2 3 4 5 6 7 8; b c d e\}} behaves 
as a self-mass
term, because the adjacent lepton propagators {\sl 1} and {\sl 9} are 
almost 
on-the-mass-shell in this limit. The reduced diagram {\sl \{1 9; a\}} then 
gives rise to a magnetic moment $M_{2^*}$, which is 
linearly IR divergent because of a two-point vertex insertion.

In the {\it K}-operation, however, 
only the UV-divergent part of the mass renormalization term is subtracted.
This is why (\ref{aX253UVfree})
contains the unsubtracted UV-finite parts of mass renormalization terms,
such as $dm_{16}^{\rm R}$, which gives rise to a linearly IR divergent
term proportional to $M_{2^*}$.
To remove this linear IR divergence\footnote{%
The linear IR divergent terms
in $M_{2^*}$ exactly cancel out within $M_{2^*}$ itself and the analytic 
value of $M_{2^*}=1$  is finite.
The cancellation, however, does not occur in  numerical integration of
our parametric integral formula $M_{2^*}$ and it suffers from the linear IR 
divergence.}
we have only to complete the standard mass-renormalization by
subtracting also the remaining part of mass-renormalization term.
This is the procedure called {\it R}-subtraction.
For instance the operation of ${\mathbb R}_{2-8}$ on $M_{\rm X253}^{\rm R}$,
implemented on {\sc gencode}{\it N}, yields 
\begin{equation}
 {\mathbb R}_{2-8} M_{\rm X253}^{\rm R}  =  M_{2^*} dm_{16}^{\rm R} ,
\label{aX253linIR}
\end{equation}
where 
$dm_{16}^{\rm R}$ is defined in (\ref{otherUVterms}).

Once all linear IR divergences are removed by {\it R}-subtractions
we are left with logarithmic IR divergences.
When the self-energy-like subdiagram $\mathcal{S}$ behaves 
as a magnetic moment amplitude of lower order and can be mimicked 
by a point (vector) vertex,
the {\it outer} residual diagram $\mathcal{R} = \mathcal{G}/\mathcal{S}$ 
behaves like a vertex diagram and its IR behavior
is exactly the same as  that of the 
vertex renormalization constant extracted from $\mathcal{R}$.
We find several residual diagrams: 
{\sl \{1 9; a\}} for the residual diagram of $\mathcal{S}_1$, 
{\sl \{1 2 3 4 8 9; a b c\}} for $\mathcal{S}_2$,
{\sl \{1 2 4 5 6 7 8 9; a b d e\}} for $\mathcal{S}_3$, 
as well as the combinations of 
{\sl \{1 9; a\}}
with the other two.

For $\mathcal{R}$ = {\sl \{1 9; a\}},
the IR divergence can be extracted by the {\it I}-subtraction
\begin{equation}
   {\mathbb I}_{19}M_{\rm X253}^{\rm R} = L_2^{\rm R} M_{16}^{\rm R} ,
\label{Iop1}
\end{equation} 
where $L_2^{\rm R}$,
which is logarithmically IR divergent,
is the UV-finite part of the second-order vertex renormalization constant $L_2$,
and $M_{16}^{\rm R}$ is the UV-divergence-free part of the
eighth-order magnetic moment $M_{16}$.

In addition the {\it I}-subtraction works on
the linearly IR-divergent terms such as ${\mathbb R}_{567} M_{\rm X253}^{\rm R}$.
The IR divergence subtraction scheme in {\sc gencode}{\it N} will give rise to
\begin{equation}
   {\mathbb I}_{19} {\mathbb R}_{567} M_{\rm X253}^{\rm R} 
      = dm_{4a}^{\rm R} L_2^{\rm R} M_{4b(1^*)}^{\rm R} ,
\label{Iop6}
\end{equation} 
\begin{equation}
   {\mathbb I}_{12489} {\mathbb R}_{567} M_{\rm X253}^{\rm R} 
      = dm_{4a}^{\rm R} L_{4b2(2^*)}^{\rm R} M_2,
\label{Iop7}
\end{equation} 
\begin{equation}
   {\mathbb I}_{19} {\mathbb I}_{248} {\mathbb R}_{567} M_{\rm X253}^{\rm R} 
      = dm_{4a}^{\rm R} L_2^{\rm R} L_{2^*}^{\rm R} M_2.
\label{Iop8}
\end{equation} 
It is easy to check that (\ref{Iop6}) gives a correct IR-divergent term 
as expected. 
It turns out, however, that the prescription encoded in {\sc gencode}{\it N} 
for the construction of the subtraction terms (\ref{Iop7}) and (\ref{Iop8}) 
has some discrepancy from the formulation 
that stems from the choice in the separation of the finite and divergent 
parts of the term $L^{*}$, 
and it actually induces IR divergence in (\ref{Iop8}). 

To understand the reason for this let us recall 
the order of IR divergence 
${\mathbb I}_{19} {\mathbb I}_{248} {\mathbb R}_{567} M_{\rm X253}^{\rm R}$.
The IR divergence associated with $\mathcal{S}_1$
(which corresponds to ${\mathbb I}_{19}$) 
is a necessary condition of 
the IR divergence associated with $\mathcal{S}_3$
(which corresponds to ${\mathbb I}_{248}$), 
since the reduced subdiagram {\sl \{1 9; a\}} is \textit{included} 
in the reduced subdiagram {\sl \{1 2 4 5 6 7 8 9; a b d e\}}.
Similarly, the simultaneous IR divergence of ${\mathbb I}_{19}$ and 
${\mathbb I}_{248}$ is the necessary condition of the self-mass term of 
$dm_{4a}^{\rm R}$.
This suggests that the diagram ${\rm X253}$ should have the IR divergence
of the form 
\begin{equation}
  L_2^{\rm R}(\text{\sl 1,9})\,
  L_{2^*}(\text{\sl 2,4,8})\,
  dm_{4a}(\text{\sl 5,6,7})\,
  M_2(\text{\sl 3}),
\end{equation}
where we indicate the lepton lines consisting of each term 
in the parentheses.
The mass-renormalization term $dm_{4a}(\text{\sl 5,6,7})$ can be exactly removed
by {\it K}-operation and {\it R}-subtraction as described before.

The {\it I}-operation encoded in {\sc gencode}{\it N} creates
an IR subtraction term of the form
\begin{equation}
L^{\rm R} = L - L^{\rm UV} - \text{UV divergences of subdiagrams}
\label{Iop11}
\end{equation} 
for the vertex renormalization constant $L$. 
(See (\ref{LR}) for the precise definition.)
By the construction of {\it K}-operation, $L^{\rm UV}$ is identified 
as the {\it maximally-contracted} term. (See Ref.~\cite{Aoyama:2005kf}.)
When the {\it I}-subtraction is accompanied by the {\it R}-subtraction 
from inner part of the diagram, it yields the term of the form $L^*$ 
where ${*}$ stands for the insertion of a two-point vertex in one of 
the lepton lines of $L$. 
{\sc gencode}{\it N} ignores this difference in the IR-subtraction step 
and applies the same rule to $L^{*}$ and $L$, 
constructing the IR subtraction term of the form
\begin{equation}
  \tilde{L}^{* \rm R} 
  = L^{*} - L^{*}\bigr|_\text{max.~contr.} 
    - \text{UV divergences of subdiagrams}.
\label{Iop12}
\end{equation} 
Note that 
$\tilde{L}^{* \rm R}$ is distinguished from 
$L^{* \rm R} = L^{*} - \text{(UV divergences of subdiagrams)}$, 
where 
$L^{*}$ does not suffer from an overall UV divergence 
as is easily seen by UV power counting. 
We may use $\tilde{L}^{* \rm R}$ instead of $L^{* \rm R}$ 
in order to subtract IR divergence. 
The difference simply results in the additional and finite residual 
renormalization terms proportional to $\Delta L^{*}\ \Delta dm$ that have been 
correctly incorporated in our calculation. 

For some specific diagrams in which the structure $L^{*}$ appears 
inside of another IR-divergent structure $L^{R}$, 
the finite contribution of $L^{*}\bigr|_\text{max.~contr.}$ 
induces spurious IR divergence. 
To see this, 
let us go back to our case of {\rm X253} and express the
{\it contraction structure} of $L_{4b2(2^*)}$ in Eq.~(\ref{Iop7})
symbolically as $L_{4b2(2^*)} \equiv F_0 + F_1 + F_2$, 
where $F_i$ is the term with $i$ contractions.
$F_2$ corresponds to $L_{4b2(2^*)}\bigr|_\text{max.~contr.}$
and $F_0 + F_1$ corresponds to $\tilde{L}_{4b2(2^*)}$.
The IR-divergence structure of $L_{4b2(2^*)}$ can be isolated
by the ${\mathbb I}_{19}$ operation as
\begin{equation}
{\mathbb I}_{19} L_{4b2(2^*)}^{\rm R} = L_2^{\rm R} L_{2^*}.
\label{Iop9}
\end{equation} 
After extraction of $L_2^{\rm R}$ by the ${\mathbb I}_{19}$ operation, 
the remaining factor has the contraction structure
$F_0 + F_1$, where $F_0$ and $F_1$ correspond to $\tilde{L}_{2^*}^{\rm R}$ 
and $\Delta L_{2^*}$, respectively.
Substituting (\ref{Iop9}) in (\ref{Iop7}) one finds that the result is
different from (\ref{Iop8}) by
\begin{equation}
       L_2^{\rm R}\ \Delta L_{2^*}\ dm_{4a}^{\rm R}\ M_2,
\label{Iop10}
\end{equation} 
which is logarithmically IR-divergent due to the presence of $L_2^{\rm R}$.
Since $L_{2^*}$ is UV-finite, 
there is no $L_{2^*}^{\rm UV}$ to be subtracted by 
the {\it K}-operation.
Thus the {\it I}-operation as defined in
{\sc gencode}{\it N} yields spurious IR divergence for $M_{\rm X253}$. 
At present this is corrected by adding (\ref{Iop10}) to (\ref{Iop8}) 
manually. 
This modification had been adopted in the calculation presented in 
Ref.~\cite{ae10:PRL}. 

Note that the spurious divergent term in {\sc gencode}{\it N} 
emerges first at tenth order. 
It occurs when there are nested {\it I}-operations and 
the inner part also involves self-mass subtraction.
Since {\it R}-subtractions are applied to fourth or higher order 
self-energy-like subdiagrams, the total order of diagram 
should be at least ten. 
There are only two diagrams in tenth order, $X253$ and $X256$. 

To summarize, the IR divergences of $M_{\rm X253}^{\rm R}$ can be separated 
by considering all combination of {\it R}- and {\it I}-subtractions. 
After separating IR-divergent and IR-finite parts of other 
terms of (\ref{aX253UVfree}) in the same fashion, 
we obtain:
\begin{align}
 a_{\rm X253} &=  \Delta \!M_{\rm X253}
       + \Delta \!M_{16} ~ L_2^{\rm R} 
       - \Delta \!M_{42} ~ B_2^{\rm R} 
       - 2 \Delta \!M_{30} ~ L_2^{\rm R} 
\nonumber \\
       &- 2 \Delta \!M_{6a} ~ (L_2^{\rm R})^2 
       + \Delta \!M_{6b} ~ ( 4~L_2^{\rm R}~B_2^{\rm R} - B_{4a}^{\rm R} )
       - \Delta \!M_{6c} ~ L_2^{\rm R}~B_2^{\rm R} 
\nonumber \\
       &+ \Delta \!M_{4a} ~ \{  
                              - B_2^{\rm R}~L_{4b2}^{\rm R} 
                             + L_{6b(2)}^{\rm R}  \} 
\nonumber \\
       &+ \Delta \!M_{4b} ~ \{  - 2~L_2^{\rm R}~(B_2^{\rm R})^2 
                            + 4~(L_2^{\rm R})^2~B_2^{\rm R} 
                            + B_{4a}^{\rm R}~(B_2^{\rm R} - L_2^{\rm R} ) \}
\nonumber \\
       &+ M_2 ~ dm_{4a}^{\rm R} ~ (  - \tilde{L}_{4b2(2^*)}^{\rm R} + B_{4b(1^*)}^{\rm R} - B_{2^*}~B_2^{\rm R} )
\nonumber \\
       &+ M_2 ~ [  L_{42(2)}^{\rm R} - B_{16}^{\rm R} 
               +  B_{6c}^{\rm R}~B_2^{\rm R} 
               +  2~B_{6a}^{\rm R}~L_2^{\rm R} 
               -  4~L_{6b(2)}^{\rm R}~L_2^{\rm R} 
\nonumber \\
       &\quad  + 4~L_{4b2}^{\rm R}~L_2^{\rm R}~B_2^{\rm R} 
               + B_{4a}^{\rm R} ~ \{ B_{4b}^{\rm R} - L_{4b2}^{\rm R}
               - (B_2^{\rm R})^2 + L_2^{\rm R}~B_2^{\rm R} \}
\nonumber \\
       &\quad  - 4~B_{4b}^{\rm R}~L_2^{\rm R}~B_2^{\rm R} 
               + 2~L_2^{\rm R}~(B_2^{\rm R})^2 (B_2^{\rm R} - L_2^{\rm R}) 
               ]~ .
\end{align}

\section{Summing up Residual Renormalization Terms of Set~V}
\label{sec:appendixRes}

\subsection{Preliminary remarks}

The total number of residual renormalization terms 
contributing to the Set~V of the tenth-order $g-2$ exceeds 11,000.
Evaluating these integrals individually and then combining them into one
could become intractable unless they are organized systematically.
Fortunately, it is possible to express them in terms of
lower-order $g-2$ and finite parts of lower-order renormalization
constants.
In this Appendix we will present our result following the pattern
described for lower-order cases in Appendix~A of Ref.~\cite{Aoyama:2011dy}. 

Throughout this article we are concerned  only with 
the diagrams of {\it q}-type, namely diagrams without closed lepton loops.
$M_{n}$, $n=2$, $4$,~$\cdots$, refers to 
the magnetic moment projection of
the sum of the set of unrenormalized vertex amplitudes 
transformed by means of the Ward-Takahashi identity (\ref{WTderived}), 
given in the form
\begin{align}
  M_{10}  &= \sum_{\alpha = 001}^{389} \eta_\alpha M_\alpha,
&
  M_{8}   &= \sum_{\alpha = 01}^{47} \eta_\alpha M_\alpha,
\nonumber \\
  M_{6}   &= \sum_{\alpha = A}^{H} \eta_\alpha M_\alpha,
&
  M_{4}   &= M_{4a} + M_{4b},
\label{Mn}
\end{align}
where $\eta_{\alpha} =1 $ for the time-reversal-symmetric diagrams
and $\eta_{\alpha} =2$ otherwise.
Quantities such as $L_n$, $B_n$, and $dm_n$ refer to the on-shell
renormalization constants of
vertex, wave-function, and mass renormalization types.
The quantity $L_{2^*}$ means a diagram derived from $L_2$
by insertion of a two-point vertex in the lepton line.
$L_{4^*}$ represents the set of diagrams obtained by insertion of
a two-point vertex in the lepton lines of $L_4$ in all possible ways.
$M_{n^*}$, $L_{n^*}$, $B_{n^*}$, and $dm_{n^*}$ are defined similarly.
$M_{n^{**}}$, $L_{n^{**}}$, $B_{n^{**}}$, and $dm_{n^{**}}$ are
insertion of two two-point vertices in the lepton lines of
$M_{n}$, $L_{n}$, $B_{n}$, and $dm_{n}$, and so on.
 
The UV-divergent part of quantities defined by the {\it K}-operation
is identified by the superscript $UV$.
Quantities with the superscript $R$ are the UV-finite parts
that remain after all UV-divergent parts,
including UV divergences of subdiagrams, are subtracted out.
Symbols with  prefix $\Delta$ mean UV- and IR-finite quantities.

In order to make the process of residual renormalization transparent
it is useful to treat UV-divergence subtraction, {\it R}-subtraction, 
and IR-divergence subtraction, separately, 
since {\it K}-operation and {\it I}-operation correspond
to different divergence structures.
Diagrams {\rm X253} and {\rm X256} require some modification of the {\it I}-operation.
This is discussed in Appendix~\ref{sec:appendixX253}.

\subsection{Standard on-the-mass-shell renormalization of $A_1^{(10)} [{\rm Set~V}]$}

The tenth-order magnetic moment $A_1^{(10)} [\text{Set~V}]$ 
has contributions from 389 Ward-Takahashi-summed diagrams shown in Fig.~\ref{fig:M10}. 
In the standard renormalization it
can be written in terms of unrenormalized amplitudes 
$M_{10}$, $M_8$, $M_6$, {\it etc.},
and various renormalization constants as follows:
\begin{equation}
 A_1^{(10)} [\text{Set~V}]  = \Xi_1 +\Xi_2 + \Xi_3 + \Xi_4 + \Xi_5 ,
\label{standardrenomalization}
\end{equation}
where
%
%----------------------------------------------------------------
\begingroup
%\allowdisplaybreaks
\begin{align}
 \Xi_1 
      & = M_{10}
\nonumber \\
      & + M_8  (- 7B_2 - 8L_2 )
\nonumber \\
      & + M_6  ( -5B_4 -6L_4 +20 B_2^2 +52 B_2 L_2 +33 L_2^2 ) 
\nonumber \\
      & + M_4  ( - 3B_6 - 4L_6 + 24B_4 B_2 + 32B_4 L_2 + 34B_2 L_4 + 44L_2 L_4 
\nonumber \\
      &\quad   - 28 B_2^3 -128 B_2^2 L_2 - 187B_2 L_2^2 -88 L_2^3 ) 
\nonumber \\
      & + M_2 ( - B_8 - 2L_8 + 8B_6 B_2 +12 B_6 L_2 + 16B_2 L_6 +22L_6 L_2 
\nonumber \\
      &\quad   + 4 B_4^2 - 28 B_4 B_2^2 - 96 B_4 B_2 L_2 + 14B_4 L_4 - 77 B_4 L_2^2 
\nonumber \\
      &\quad   + 14 B_2^4 +112B_2^3 L_2 -56 B_2^2 L_4 
\nonumber \\
      &\quad   + 308 B_2^2 L_2^2 - 176 B_2 L_2 L_4 + 352 B_2 L_2^3 
\nonumber \\
      &\quad   + 11 L_4^2 - 132 L_4 L_2^2 + 143 L_2^4 )
,
\label{standardrenoma10_part1}
%\nonumber \\
%
\end{align}
\begin{align}
 \Xi_2 
      & = M_{8^*} dm_2 (-1)
\nonumber \\
      & + M_{6^*} dm_2 (7 B_{2} + 8 L_{2})
\nonumber \\
      & + M_{4^*} dm_2 (5 B_{4} + 6 L_{4}-20 B_2^2 -52 B_2 L_{2} -33 L_2^2)
\nonumber \\
      & + M_{2^*} dm_2 ( 3 B_6 - 32 L_2 B_4 + 88 L_2^3 - 44 L_4 L_2 + 4 L_6 - 24 B_2 B_4 
\nonumber \\
      &\quad   + 187 B_2 L_2^2 - 34 B_2 L_4 + 128 B_2^2 L_2 + 28 B_2^3 )
\nonumber \\
      & + M_{6} dm_2 (5 B_{2^*} + 12 L_{2^*})
\nonumber \\
      & + M_{4} dm_2 (3 B_{4^*} + 4 L_{4^*}-24 B_2 B_{2^*} -68 B_2 L_{2^*}
      -32 L_2 B_{2^*} -88 L_2 L_{2^*})
\nonumber \\
      & + M_{2} dm_2 ( B_{6^{*}} +2 L_{6^{*}} -8 B_2 B_{4^*} -16 B_2 L_{4^*}
\nonumber \\
      &\quad  + 28 B_2^2 B_{2^*} +112 B_{2}^2 L_{2^*} +96 B_2 L_2 B_{2^*} 
\nonumber \\
      &\quad  + 352 B_2 L_2 L_{2^*} -12 L_2 B_{4^*} -22 L_2 L_{4^*} +77 L_2^2 B_{2^*}
\nonumber \\
      &\quad  + 264 L_2^2 L_{2^*} -8 B_{2^*} B_4 -14 B_{2^*} L_4 -28 L_{2^*} B_4
     - 44 L_{2^*} L_4 ) 
,
\label{standardrenoma10_part2}
%\nonumber \\
%
\end{align}
\begin{align}
 \Xi_3
      & = M_{6^*} dm_4 (-1)
\nonumber \\
      & + M_{6^{*}} dm_2 dm_{2^*}  
\nonumber \\
      & + M_{4^*} dm_4 ( 7 B_{2} + 8 L_{2})
\nonumber \\
      & + M_{4^*} dm_2 dm_{2^*} (-7 B_{2} -8 L_{2} ) 
\nonumber \\
      & + M_{2^*} dm_4  (5 B_4 +6 L_4 - 20 B_2^2-52 B_2 L_2 -33 L_2^2  ) 
\nonumber \\
      & + M_{2^*} dm_2 dm_{2^*} (-5 B_{4} -6 L_{4} + 20 B_2^2 + 52 B_2 L_2 +33 L_2^2 )
\nonumber \\
      & + M_4 dm_4 ( 3 B_{2^*} + 8 L_{2^*})
\nonumber \\
      & + M_{4} dm_2 dm_{2^*} (-3 B_{2^*} -8 L_{2^*} ) 
\nonumber \\
      & + M_{2} dm_{4}  ( B_{4^*} +2 L_{4^*}-8 B_2 B_{2^*} -32 B_2 L_{2^*} -12 L_2 B_{2^*} -44 L_2 L_{2^*} ) 
\nonumber \\
      & + M_{2} dm_2 dm_{2^*} (- B_{4^*} -2 L_{4^*} +8 B_2 B_{2^*} + 32 B_2 L_{2^*} + 12 L_2 B_{2^*} +44 L_2 L_{2^*} )
\nonumber \\
      & + M_{6^{**}} dm_2^2 
\nonumber \\
      & + M_{4^{**}} dm_2^2 (-7 B_{2} - 8 L_{2})
\nonumber \\
      & + M_{2^{**}} dm_2^2 (-5 B_{4} -6L_{4} +20 B_2^2 +52 B_2 L_2 + 33 L_2^2 )
\nonumber \\
      & + M_{4^*} dm_2^2 (-5 B_{2^{*}} - 12 L_{2^{*}})
\nonumber \\
      & + M_{2^*} dm_2^2 (-3 B_{4^{*}} -4L_{4^{*}} +24 B_2 B_{2^{*}} + 68 B_2 L_{2^{*}} + 32 L_2 B_{2^{*}} +88 L_2 L_{2^{*}} )
\nonumber \\
      & + M_{4} dm_2^2 (-3 B_{2^{**}} - 4 L_{2^{**}})
\nonumber \\
      & + M_{2} dm_2^2 (- B_{4^{**}} -2 L_{4^{**}} +8 B_2 B_{2^{**}} + 44 L_{2^{*}}^2
\nonumber \\
      &\quad  + 16 B_2 L_{2^{**}} + 12 L_2 B_{2^{**}} +22 L_2 L_{2^{**}} + 4 B_{2^*}^2 +28  B_{2^*} L_{2^{*}} )
,
\label{standardrenoma10_part3}
%\nonumber \\
%
\end{align}
\begin{align}
 \Xi_4
      & = M_{4^*} dm_6 (-1)
\nonumber \\
      & + M_{4^*} dm_2 dm_{4^*} 
\nonumber \\
      & + M_{4^*} dm_4 dm_{2^*}
\nonumber \\
      & + M_{4^*} dm_2 dm_{2^*}^2 (-1 ) 
\nonumber \\
      & + M_{4^{*}} dm_2^2 dm_{2^{**}} (-1)
\nonumber \\
      & + M_{2^*} dm_6  (7 B_2 +8 L_2 ) 
\nonumber \\
      & + M_{2^*} dm_2 dm_{4^*} (  - 7 B_2 - 8 L_2 )
\nonumber \\
      & + M_{2^*} dm_{4} dm_{2^*}  ( -7 B_{2} -8 L_{2} ) 
\nonumber \\
      & + M_{2^*} dm_2 dm_{2^*}^2 (7 B_{2} + 8 L_{2}) 
\nonumber \\
      & + M_{2^*} dm_2^2 dm_{2^{**}} (7 B_{2} + 8 L_{2})
\nonumber \\
      & + M_{2} dm_{6}  ( B_{2^*} +4 L_{2^*} ) 
\nonumber \\
      & + M_{2} dm_2 dm_{4^*} (- B_{2^{*}} -4 L_{2^{*}} ) 
\nonumber \\
      & + M_2 dm_4 dm_{2^*} (  - B_{2^*} - 4 L_{2^*} )
\nonumber \\
      & + M_{2} dm_2 dm_{2^*}^2 ( B_{2^*} +4 L_{2^*} ) 
\nonumber \\
      & + M_{2} dm_2^2 dm_{2^{**}} ( B_{2^{*}} +4 L_{2^{*}})
\nonumber \\
      & + M_{4^{**}} dm_2 dm_{4} (2)
\nonumber \\
      & + M_{4^{**}} dm_2^2 dm_{2^*} (-2)
\nonumber \\
      & + M_{2^{**}} dm_2 dm_4 ( - 14 B_2 - 16 L_2  )
\nonumber \\
      & + M_{2^{**}} dm_2^2 dm_{2^*} ( 14 B_{2} + 16 L_{2} )
\nonumber \\
      & + M_{2^*} dm_2 dm_{4}  (-8 B_{2^*} -20 L_{2^*} ) 
\nonumber \\
      & + M_{2^*} dm_2^2 dm_{2^*} (8 B_{2^{*}} +20 L_{2^{*}})
\nonumber \\
      & + M_{2} dm_2 dm_4 (-2 B_{2^{**}} -4 L_{2^{**}} ) 
\nonumber \\
      & + M_{2} dm_2^2 dm_{2^*} (2 B_{2^{**}} +4 L_{2^{**}})
\nonumber \\
      & + M_{4^{***}} dm_2^3 (-1)
\nonumber \\
      & + M_{2^{***}} dm_2^3 ( 7 B_{2} + 8 L_{2})
\nonumber \\
      & + M_{2^{**}} dm_2^3 ( 5 B_{2^{*}} + 12 L_{2^{*}})
\nonumber \\
      & + M_{2^*} dm_2^3 ( 3 B_{2^{**}} + 4 L_{2^{**}})
\nonumber \\
      & + M_{2} dm_2^3 (B_{2^{***}} +2 L_{2^{***}})
,
\label{standardrenoma10_part4}
%\nonumber \\
%
\end{align}
\begin{align}
 \Xi_5
      &= M_{2^*} dm_8  (-1 ) 
\nonumber \\
      & + M_{2^*} dm_2 dm_{6^*}   
\nonumber \\
      & + M_{2^*} dm_4 dm_{4^*}   
\nonumber \\
      & + M_{2^*} dm_2 dm_{2^*} dm_{4^*}  (-2 ) 
\nonumber \\
      & + M_{2^*} dm_2^2 dm_{4^{**}} (-1)
\nonumber \\
      & + M_{2^*} dm_6 dm_{2^*}
\nonumber \\
      & + M_{2^*} dm_4 dm_{2^*}^2 (  - 1 )
\nonumber \\
      & + M_{2^*} dm_2 dm_{2^*}^3  
\nonumber \\
      & + M_{2^{*}} dm_2^2 dm_{2^*} dm_{2^{**}} (3)
\nonumber \\
      & + M_{2^{*}} dm_2 dm_{2^{**}} dm_{4}  (-2 ) 
\nonumber \\
      & + M_{2^{*}} dm_2^3 dm_{2^{***}} 
\nonumber \\
      & + M_{2^{**}} dm_2 dm_6 ( 2 )
\nonumber \\
      & + M_{2^{**}} dm_2^2 dm_{4^{*}} (-2)
\nonumber \\
      & + M_{2^{**}} dm_4^2
\nonumber \\
      & + M_{2^{**}} dm_2 dm_{2^*} dm_{4}  (-4 ) 
\nonumber \\
      & + M_{2^{**}} dm_2^2 dm_{2^*}^2 (3)
\nonumber \\
      & + M_{2^{**}} dm_2^3 dm_{2^{**}} (2)
\nonumber \\
      & + M_{2^{***}} dm_2^2 dm_{4} (-3)
\nonumber \\
      & + M_{2^{***}} dm_2^3 dm_{2^*} (3)
\nonumber \\
      & + M_{2^{****}} dm_2^4 .
\label{standardrenoma10_part5}
\end{align}
\endgroup
%----------------------------------------------------------------

Terms containing self-mass subdiagrams are numerous but can be readily identified
since they always accompany some $B_n$.
For instance, $-M_{8^*} dm_2$ accompanies $-7 M_8 B_2$.

\subsection{Treatment of UV divergences by the {\it K}-operation}

Terms listed in Eqs.~%
(\ref{standardrenoma10_part1}), 
(\ref{standardrenoma10_part2}),
(\ref{standardrenoma10_part3}),
(\ref{standardrenoma10_part4}), and~%
(\ref{standardrenoma10_part5})
are all UV-divergent.
Application of {\it K}-operations to each of these
integrals extracts UV-divergent parts. 
The resulting UV-finite part will be denoted as $M_{n}^{\rm R}$, etc.
See Eqs.~(\ref{eq:Kforest}), (\ref{dmR}), and~(\ref{LR}).
{\it K}-operations applied to $M_{10}$, the first term of (\ref{standardrenoma10_part1}),
give rise to $M_{10}^{\rm R}$:
\begin{eqnarray} 
 M_{10}^R & =& M_{10}   \nonumber \\
      & +& M_8  (-7B_2^{UV} - 8L_2^{UV} ) \nonumber \\
      & +& M_6  (-5B_4^{UV} -6L_4^{UV} +20 (B_2^{UV})^2 +52 B_2^{UV} L_2^{UV} +33 (L_2^{UV})^2 ) 
\nonumber \\
      &+& M_4  ( -3B_6^{UV} - 4L_6^{UV} + 24B_4^{UV} B_2^{UV} + 32B_4^{UV} L_2^{UV} 
  + 34B_2^{UV} L_4^{UV} 
  + 44L_2^{UV} L_4^{UV} 
\nonumber \\
       & & - 28 (B_2^{UV})^3 -128 (B_2^{UV})^2 L_2^{UV} 
  - 187B_2^{UV} (L_2^{UV})^2 
  - 88 (L_2^{UV})^3 ) + \ldots,
\label{KoponA10}
\end{eqnarray} 
where the remaining terms are not shown explicitly,
but can be readily found
since the coefficients of all UV-divergent terms of (\ref{KoponA10})
are the same as those of standard renormalization formula 
(\ref{standardrenomalization}).

Solving (\ref{KoponA10}) for $M_{10}$ and
substituting these terms in 
(\ref{standardrenomalization}), 
one can express $A_1^{(10)} [\text{Set~V}]$ in terms of $M_{10}^{\rm R}$,
$M_8$, $M_6$, etc.
Next, replace $M_8$ by $M_8^{\rm R}$, etc., using Eq.~(A24) of 
Ref.~\cite{Aoyama:2011dy}.
The result still contains $M_6$, which can be replaced by $M_6^{\rm R}$
using Eq.~(A14) of Ref.~\cite{Aoyama:2011dy}, and so on.
We also have to extract UV-finite parts
of renormalization constants $L_n$, $B_n$, $dm_n$, etc.
In this way we arrive at the expression of
$A_1^{(10)} [\text{Set~V}]$ as the sum of UV-finite quantities only:
\begingroup
\allowdisplaybreaks
\begin{align}
 A_1^{(10)} [\text{Set~V}] 
      &= M_{10}^{\rm R}   
\nonumber \\
      &+ M_8^{\rm R}  (- 7B_2^{\rm R} - 8L_2^{\rm R} ) 
\nonumber \\
      &+ M_6^{\rm R}  ( 
-5B_4^{\rm R} 
-6L_4^{\rm R} 
+20 (B_2^{\rm R})^2 
+ 52 B_2^{\rm R} L_2^{\rm R} 
+33 (L_2^{\rm R})^2 
) 
\nonumber \\
      &+ M_4^{\rm R}  (
- 3B_6^{\rm R} 
- 4L_6^{\rm R} 
+ 24 B_2^{\rm R} B_4^{\rm R} 
+ 32L_2^{\rm R} B_4^{\rm R} 
+ 34B_2^{\rm R} L_4^{\rm R} 
+ 44 L_2^{\rm R} L_4^{\rm R} 
\nonumber \\
      &\quad  - 28 (B_2^{\rm R})^3 
-128(B_2^{\rm R})^2 L_2^{\rm R} 
-187 B_2^{\rm R} (L_2^{\rm R})^2 
-88 (L_2^{\rm R})^3 
) 
\nonumber \\
      &+ M_2 (
- B_8^{\rm R} 
- 2L_8^{\rm R} 
+ 8 B_6^{\rm R} B_2^{\rm R} 
+ 12L_2^{\rm R} B_6^{\rm R} 
+ 16 B_2^{\rm R} L_6^{\rm R} 
+ 22 L_2^{\rm R} L_6^{\rm R} 
\nonumber \\
      &\quad + 4 (B_4^{\rm R})^2 
- 28 (B_2^{\rm R})^2 B_4^{\rm R} 
- 96 B_2^{\rm R} L_2^{\rm R} B_4^{\rm R} 
+ 14 L_4^{\rm R} B_4^{\rm R}  
- 77  (L_2^{\rm R})^2 B_4^{\rm R} 
\nonumber \\
      &\quad + 14 (B_2^{\rm R})^4
+112(B_2^{\rm R})^3 L_2^{\rm R} 
- 56 (B_2^{\rm R})^2 L_4^{\rm R} 
\nonumber \\
      &\quad + 308 (B_2^{\rm R})^2 (L_2^{\rm R})^2 
  -  176B_2^{\rm R} L_2^{\rm R} L_4^{\rm R} 
+ 352 (L_2^{\rm R})^3 B_2^{\rm R} 
\nonumber \\
      &\quad + 11 (L_4^{\rm R})^2 
- 132 (L_2^{\rm R})^2 L_4^{\rm R} 
+ 143 (L_2^{\rm R})^4 
)
\nonumber \\
      &+ M_4^{\rm R} dm_4^{\rm R} (3 B_{2^*} + 8 L_{2^*})
\nonumber \\
      &+ M_2 dm_4^{\rm R} ( B_{4^*}^{\rm R} + 2 L_{4^*}^{\rm R} )
\nonumber \\
      &+ M_2 dm_4^{\rm R} (-12 B_{2^*} L_2^{\rm R}-  8 B_2^{\rm R} B_{2^*} 
        - 44 L_{2^*}  L_2^{\rm R} -32 B_2^{\rm R} L_{2^*} )
\nonumber \\
      &+ M_2 dm_6^{\rm R} (B_{2^*} +4 L_{2^*} )
\nonumber \\
      &+ M_2 dm_4^{\rm R}   dm_{2^*}^{\rm R} (- B_{2^*} -4 L_{2^*} )
\nonumber \\
      &+ M_{6^*}^{\rm R} (- dm_4^{\rm R})
\nonumber \\
      &+ M_{4^*}^{\rm R} dm_4^{\rm R} (7 B_2^{\rm R} + 8 L_2^{\rm R})
\nonumber \\
      &+ M_{2^*} dm_4^{\rm R} ( -52 L_2^{\rm R} B_2^{\rm R} 
        - 33 (L_2^{\rm R})^2 - 20 (B_2^{\rm R})^2 
        + 5 B_4^{\rm R}  + 6 L_4^{\rm R} )
\nonumber \\
      &+ M_{4^*}^{\rm R} (-dm_6^{\rm R} 
        + dm_4^{\rm R} dm_{2^*}^{\rm R})
\nonumber \\
      &+ M_{2^*} dm_6^{\rm R} (7 B_2^{\rm R} + 8 L_2^{\rm R})
\nonumber \\
      &+ M_{2^*} dm_4^{\rm R} dm_{2^*}^{\rm R} (-7 B_2^{\rm R} -8 L_2^{\rm R})
\nonumber  \\
      &+ M_{2^*} ( - dm_8^{\rm R} )
\nonumber \\
      &+ M_{2^*} (dm_4^{\rm R} dm_{4^*}^{\rm R}  
        + dm_{2^*}^{\rm R} dm_6^{\rm R} - (dm_{2^*}^{\rm R} )^2 dm_4^{\rm R})
\nonumber \\
      &+ M_{2^{**}} (dm_4^{\rm R} )^2.
\label{a10intermediate}
\end{align}
\endgroup

Note that Eq.~(\ref{a10intermediate}) has exactly the same structure as
(\ref{standardrenomalization}).
Apparent dramatic simplification of
(\ref{a10intermediate}) is a consequence of the fact that
$dm_2^{\rm R} $ vanishes according to the definition of $K$-operation.
This is what one would expect since
all UV-divergent quantities 
in (\ref{standardrenomalization}) must cancel out
after $K$-operation is carried out, leaving only UV-finite
pieces with their original numerical coefficients unchanged.

\subsection{{\it R}-subtraction}

Eight of the last nine terms of Eq.~(\ref{a10intermediate}) containing
$M_{2^*}$, $M_{4^*}^{\rm R}$, and $M_{6^*}^{\rm R}$, 
are linearly IR-divergent.
The last term proportional to $M_{2^{**}}$
is even more singular, being quadratically IR-divergent.
They are all characterized by the fact that they contain one of the factors
$dm_4^{\rm R}$, $dm_6^{\rm R}$, or $dm_8^{\rm R}$,
which are UV-finite remnants of $dm_4$, $dm_6$, or $dm_8$,
after their UV-divergent parts are removed by the {\it K}-operation.
Since $ A_1^{(10)} [\text{Set~V}]$  as a whole is IR-finite, these IR-divergent terms
must cancel linear or quadratic IR divergences hidden in $M_8^{\rm R}$
and $M_{10}^{\rm R}$.
The {\it R}-subtraction is a procedure to combine and cancel out
corresponding IR divergences of $M_8^{\rm R}$ or $M_{10}^{\rm R}$
with those of the last nine terms of
(\ref{a10intermediate}),
which amounts to redefining $M_{8}^{\rm R}$ and  $M_{10}^{\rm R}$.
The last nine terms of
(\ref{a10intermediate}) must be dropped after 
$M_{8}^{\rm R}$ and  $M_{10}^{\rm R}$ are redefined.
This procedure is incorporated in {\sc gencode}{\it N}
as its integral part.

\subsection{Separation of IR divergences by {\it I}-operation}

After linear IR divergences are removed by the {\it R}-subtraction
we still have to deal with logarithmic IR divergences.
This can be readily handled by the {\it I}-operation.
However, the {\it I}-operation incorporated in
the program {\sc gencode}{\it N} requires a slight modification
for the diagrams {\rm X253} and {\rm X256}, which is described in 
Appendix~\ref{sec:appendixX253}.

The result of the {\it I}-operation can be factorized analytically
into the product of UV-finite parts of lower-order 
renormalization constant and anomalous magnetic moment
as is shown in Eq.~(\ref{IoponMG}).
The individual UV-finite terms of (\ref{a10intermediate}) 
are expressed as sums of IR-divergent parts and IR-finite parts
(which are indicated by the prefix $\Delta$). 
The sums of the finite magnetic moment amplitudes of the $n$th-order 
are given in terms of UV-finite quantities as follows:
\begingroup
\allowdisplaybreaks
\begin{align}
\Delta M_{10} 
  &= M_{10}^{\rm R}
   - M_{8}^{\rm R} L_2^{\rm R}
   - M_{6}^{\rm R} ( L_4^{\rm R} - (L_2^{\rm R})^2 )
\nonumber \\
  &- M_{4}^{\rm R} ( L_6^{\rm R} - 2 L_2^{\rm R} L_4^{\rm R} + (L_2^{\rm R})^3 - 2 \tilde{L}_{2^*}^{\rm R} dm_4^{\rm R} )
\nonumber \\
  &- M_{2} ( L_8^{\rm R} - 2 L_2^{\rm R} L_6^{\rm R} - (L_4^{\rm R})^2 + 3 (L_2^{\rm R})^2 L_4^{\rm R} - (L_2^{\rm R})^4
\nonumber \\
  &\quad 
  - 2 \tilde{L}_{2^*}^{\rm R} dm_6^{\rm R} 
  + 2 \tilde{L}_{2^*}^{\rm R} dm_{2^*}^{\rm R} dm_4^{\rm R}
  + 2 \tilde{L}_{2^*}^{\rm R} L_2^{\rm R} dm_4^{\rm R}
  + 2 L_2^{\rm R} L_{2*} dm_4^{\rm R}
  - \tilde{L}_{4^*}^{\rm R} dm_4^{\rm R} )
\nonumber \\
  &- M_{6^*}^{\rm R} dm_4^{\rm R}
\nonumber \\
  &- M_{4^*}^{\rm R} ( dm_6^{\rm R} - dm_4^{\rm R} L_2^{\rm R} - dm_{2^*}^{\rm R} dm_4^{\rm R} )
\nonumber \\
  &- M_{2^*} ( 
  dm_8^{\rm R} 
  - dm_{4^*}^{\rm R} dm_4^{\rm R}
  + (dm_{2^*}^{\rm R})^2 dm_4^{\rm R}
  - dm_{2^*}^{\rm R} dm_6^{\rm R}
\nonumber \\
  &\quad
  - dm_6^{\rm R} L_2^{\rm R}
  + dm_{2^*}^{\rm R} dm_4^{\rm R} L_2^{\rm R}
  - dm_4^{\rm R} L_4^{\rm R}
  + dm_4^{\rm R} (L_2^{\rm R})^2
)
\nonumber \\
  &+ M_{2^{**}} ( dm_4^{\rm R} )^2 ,
\label{expr:DM10}
%
%\end{align}
\\
%\begin{align}
%
\Delta M_{8} 
  &= M_{8}^{\rm R}
   - M_{6}^{\rm R} L_2^{\rm R}
   - M_{4}^{\rm R} ( L_4^{\rm R} - ( L_2^{\rm R} )^2 )
\nonumber \\
  &- M_{2}^{\rm R} ( L_6^{\rm R} - 2 L_4^{\rm R} L_2^{\rm R} + (L_2^{\rm R})^3 - 2 \tilde{L}_{2^*}^{\rm R} dm_4^{\rm R})
\nonumber \\
  &- M_{2^*} ( dm_6^{\rm R} - dm_{2^*}^{\rm R} dm_4^{\rm R} - dm_4^{\rm R} L_2^{\rm R} )
\nonumber \\
  &- M_{4^*}^{\rm R} dm_4^{\rm R} ,
\label{expr:DM8}
%
%\end{align}
\\
%\begin{align}
%
\Delta M_{6} 
  &= M_{6}^{\rm R} - M_{4}^{\rm R} L_2^{\rm R} - M_2 (L_4^{\rm R} - (L_2^{\rm R})^2) - M_{2^*} dm_4^{\rm R} ,
\label{expr:DM6}
%
%\end{align}
\\
%\begin{align}
%
\Delta M_{4} 
  &= M_{4}^{\rm R} - M_{2} L_2^{\rm R} .
\label{expr:DM4}
\end{align}
\endgroup
The finite integrals derived from the renormalization constants are:
\begin{align}
  \DLB_{8} & = L_{8}^{\rm R} + B_{8}^{\rm R}  
               - \{ L_{6}^{\rm R}-2 L_{4}^{\rm R}L_2^{\rm R}+ (L_2^{\rm R})^3  \} 
                                     ( L_2^{\rm R} + B_2^{\rm R} )              
\nonumber \\
           &   - \{ L_{4}^{\rm R}- (L_2^{\rm R})^2 \} 
                                     ( L_4^{\rm R} + B_4^{\rm R} ) 
               -   L_{2}^{\rm R}    ( L_6^{\rm R} + B_6^{\rm R} ) 
\nonumber \\
           &   - \{ dm_6^{\rm R} - (L_2^{\rm R} +dm_{2^*}^{\rm R}) dm_4^{\rm R}  \}
                                     ( 2 L_{2^*} + B_{2^*} )
\nonumber \\
           &   + 2 \tilde{L}_{2^*}^{\rm R} dm_4^{\rm R} 
                                     ( L_2^{\rm R} + B_2^{\rm R} ) 
               - dm_4^{\rm R} ( L_{4^*}^{\rm R} + B_{4^*}^{\rm R} )
,
\nonumber \\
  \DLB_{6} & = L_{6}^{\rm R} + B_{6}^{\rm R} 
               - \{ L_{4}^{\rm R}- (L_2^{\rm R})^2 \} ( L_2^{\rm R} + B_2^{\rm R} )
               - L_{2}^{\rm R} ( L_4^{\rm R} + B_4^{\rm R} ) 
\nonumber \\
           &   - dm_4^{\rm R} ( 2 L_{2^*} + B_{2^*} ) ,
\nonumber \\
  \DLB_{4} & = L_{4}^{\rm R} + B_{4}^{\rm R} - L_{2}^{\rm R} (L_{2}^{\rm R} + B_{2}^{\rm R} ),
\nonumber \\
  \DLB_{2} & = L_2^{\rm R} + B_2^{\rm R},
\nonumber \\
  \Delta dm_{6} & = dm_6^{\rm R} - L_{2}^{\rm R} dm_4^{\rm R},
\nonumber \\
  \Delta dm_{4} & =  dm_4^{\rm R} ~.
\label{DeltaLBdm}
\end{align}
(See Appendix~\ref{sec:appendixX253} for the quantities 
$\tilde{L}_{2^*}^{\rm R}$ and $\tilde{L}_{4^*}^{\rm R}$. )

Substituting Eqs.~(\ref{expr:DM10})--(\ref{DeltaLBdm}) in (\ref{a10intermediate}),
we can transform $A_1^{(10)} [\text{Set~V}]$ into the sum
of terms which are completely free from UV- and IR-divergences:
\begin{eqnarray}
 A_1^{(10)} [\text{Set~V}] & =& \Delta M_{10}   \nonumber \\
       &+& \Delta M_8  (  - 7 \DLB_2 ) \nonumber \\
       &+& \Delta M_6  ( 20 (\DLB_2)^2 - 5 \DLB_4 ) \nonumber \\
       &+& \Delta M_4  ( 24 \DLB_2 \DLB_4 - 28 (\DLB_2)^3 - 3 \DLB_6 ) \nonumber \\
       &+& M_2 ( 8\DLB_2 \DLB_6 - 28 (\DLB_2)^2 \DLB_4) \nonumber \\ 
       &+& M_2 ( 14 (\DLB_2)^4 + 4 (\DLB_4)^2 - \DLB_8 ) \nonumber \\
       &+& 2 \Delta M_4 \Delta L_{2^*} \Delta dm_4 \nonumber \\
       &+& 2 M_2 \Delta L_{2^*} \Delta dm_6 \nonumber \\
       &+&   M_2 \Delta L_{4^*} \Delta dm_4 \nonumber \\
       &-& 16 M_2 \Delta L_{2^*} \DLB_2 \Delta dm_4 \nonumber  \\ 
       &-& 2 M_2 \Delta L_{2^*} \Delta dm_{2^*} \Delta dm_4,
\label{residual}
\end{eqnarray}
where 
$\Delta L_{4^*} = L_{4^*}^{\rm R} - \tilde{L}_{4^*}^{\rm R}$, 
and
$\Delta L_{2^*} = L_{2^*} - \tilde{L}_{2^*}^{\rm R}$. 
The values of $\Delta L_{2^*}$, $\Delta L_{4^*}$, $\Delta dm_{6}$,
and $\Delta dm_{2^*}$ are listed in Table~\ref{Table:residual_const}.

Note that the last five terms of (\ref{residual}), 
even though they contain factors such as $\Delta dm_4$ and $\Delta dm_6$,
are not removed by the {\it R}-subtraction. 
This is because 
the factors $\Delta L_{2^*}$ and $\Delta L_{4^*}$ are not IR-divergent
so that the {\it R}-subtraction rule does not apply to them.
As a matter of fact they are indefinite, although finite,
since they depend on how the IR-divergent parts $I_{2^*}$ and $I_{4^*}$
are defined.  However, this does not cause difficulty
since these terms must be canceled by
the corresponding terms hidden in $\Delta M_{10}$ and $\Delta M_8$.
Actually this is an artifact caused by our definition of 
{\it R}-subtraction and {\it I}-operation
adopted in the program {\sc gencode}{\it N}
which subtracts only the IR-divergent parts $I_{2^*}$ and $I_{4^*}$ 
instead of full $L_{2^*}$ and $L_{4^*}$.
The value of $A_1^{(10)} [\text{Set~V}]$ is unambiguous as far as the {\it I}-operation
is carried out consistently throughout the calculation.

%================================================================
\bibliographystyle{apsrev}
\bibliography{b}

%================================================================
\end{document}